\documentclass{article}
\usepackage{amssymb,ulem}
\usepackage[colorlinks=true, pdfstartview=FitV, linkcolor=black, citecolor=black, urlcolor=blue]{hyperref}

\usepackage{tikz, mathrsfs}
\usepackage{ tikz, pgfplots, todonotes,ulem,amssymb}
\usetikzlibrary{decorations.markings,arrows, calc, shapes, intersections, patterns}

\def \scr{\mathscr}

\def\red#1{\textcolor[rgb]{0.9, 0, 0}{#1} }
\def\blue#1{\textcolor[rgb]{0,0,1.0}{#1}}
\newcommand {\TriangleCut}[2]{ %#1 in {0,1,2,3}  #4 = 0,1 for labels 
     \foreach [count=\j] \vv in {1,1,1}{  
     \draw[decoration={aspect=0.5, segment length=02, amplitude=0.7,coil},decorate] (0,0) to (-180 +120*\j-120 :1);
     \draw (-180 +120*\j-120 :0.49) 
             to  node [pos=0.6,above] {\small $J$}  (-180 +120*\j-120:0.51);}

\draw [
    postaction={decorate,decoration={markings,mark=at position 0.11 with {\arrow[line width=1pt]{<}}}},
    postaction={decorate,decoration={markings,mark=at position 0.247 with {\arrow[line width=1pt]{>}}}},
    postaction={decorate,decoration={markings,mark=at position 0.44 with {\arrow[line width=1pt]{<}}}},
    postaction={decorate,decoration={markings,mark=at position 0.584 with {\arrow[line width=1pt]{>}}}},
    postaction={decorate,decoration={markings,mark=at position 0.904 with {\arrow[line width=1pt]{>}}}},
    postaction={decorate,decoration={markings,mark=at position 0.77 with {\arrow[line width=1pt]{<}}}},
    ] (0:2) to  (120:2) to  (-120:2) to  cycle ;
       \foreach  [count=\j]  \vv in {1,1,1} {
        \draw [line width=0.4pt, 
        blue,postaction={decorate,decoration={markings,mark=at position 0.55 with {\arrow[line width=1pt]{<}}}}] (0,0) to  node[pos=0.4, above]   
           {\small $L $} (120*\j-120:2);
        \draw [fill](120*\j-120:2) circle[radius=0.02];
      }
}

\newcommand {\Triangleplain}[2]{ %#1 in {0,1,2,3}  #4 = 0,1 for labels
  
\ifthenelse{#1=0}{
     \foreach [count=\j] \vv in {0,0,0}{ 
          \ifthenelse{\vv=1}
          { \draw [decoration={aspect=0.5, segment length=02, amplitude=0.7,coil},decorate]  (0,0) 
             to  node [pos=0.6,above,sloped] {\ifthenelse{#2=1}{$i\sigma_2$}{}} (-180 +120*\j-120:1);}
          {\draw [decoration={aspect=0.5, segment length=02, amplitude=0.7,coil},decorate] 
              (0,0) to  node [pos=0.6,above,sloped] {\ifthenelse{#2=1}{$i\sigma_2$}{}} (-180 +120*\j-120 :1);}
       }
      }{}
  \ifthenelse{#2=0} {% We want the shading....
    \fill  [red!30!white, opacity=0.5 ] (60:1) to (0,0) to (0:2) to cycle;
    \fill [red!30!white, opacity=0.5 ] (-60:1) to (0,0) to (-120:2) to cycle;
    \fill [red!30!white, opacity=0.5 ] (120:2) to (0,0) to (-180:1) to cycle;
  }{}
     \draw [
    postaction={decorate,decoration={markings,mark=at position 0.106 with {\arrow[line width=1.5pt]{<}}}},
    postaction={decorate,decoration={markings,mark=at position 0.247 with {\arrow[line width=1.5pt]{>}}}},
    postaction={decorate,decoration={markings,mark=at position 0.436 with {\arrow[ line width=1.5pt]{<}}}},
    postaction={decorate,decoration={markings,mark=at position 0.5765 with {\arrow[ line width=1.5pt]{>}}}},
    postaction={decorate,decoration={markings,mark=at position 0.91 with {\arrow[line width=1.5pt]{>}}}},
    postaction={decorate,decoration={markings,mark=at position 0.77 with {\arrow[line width=1.5pt]{<}}}},
    ] (0:2) to  (120:2) to  (-120:2) to  cycle ;
       \foreach  [count=\j]  \vv in {1,1,1} {
        \ifthenelse{\vv=1}{
        \draw [line width=0.4pt, 
        blue,postaction={decorate,decoration={markings,mark=at position 0.55 with {\arrow[line width=1.2pt]{<}}}}] (0,0) to  node[pos=0.4, above,sloped]   
           {\ifthenelse{#2=1}{\tiny $L $}{}} (120*\j-120:2);
        \draw [fill](120*\j-120:2) circle[radius=0.02];
        }
}}

\def\FG{\rho}

\def\Hcal{{\mathcal H}}
\def\Fcal{{\mathcal F}}
\def\Gcal{{\mathcal G}}
\def\at{\tilde{a}}
\def\bt{\tilde{b}}
\def\Ct{\tilde{\CC}}
\def\Qcal{{\mathcal Q}}

\def\red#1{\textcolor[rgb]{0.9, 0, 0}{#1} }
\def\blue#1{\textcolor[rgb]{0, 0, 0.9}{#1} }
\def\e{\hbar}
\def\Qcal{{\mathcal Q}}
\def\Pcal{{\mathcal P}}
\def\Qcal{{\mathcal Q}}
\def \CC{{\mathcal C}}
\newlength{\dinwidth}
\def\le{\left}
\def \QED{\hfill $\blacksquare$\par \vskip 4pt}
\def\ri{\right}

\newlength{\dinmargin}
\def\bea{\begin{eqnarray}}
\def\eea{\end{eqnarray}}   
\def \wt{ \widetilde }
\def \&{\hspace{-15pt}&}
\def \d{{\mathrm d}}

\newtheorem{definition}{Definition}[section]
\newtheorem{theorem}{Theorem}[section]
\newtheorem{proposition}{Proposition}[section]
\newtheorem{corollary}{Corollary}[section]
\newtheorem{remark}{Remark}[section]
\newtheorem{lemma}{Lemma}[section]
\newtheorem{example}{Example}[section]

\def \L {\mathcal L}

\def\be{\begin{equation}}
\def\ee{\end{equation}}
\def\ben{\begin{displaymath}}
\def\een{\end{displaymath}}
\def\baa{\begin{eqnarray}}
\def\eaa{\end{eqnarray}}

\def\Gh{\widehat{G}}

\def\ba{\begin{array}}
\def\ea{\end{array}}
\makeatletter
\@addtoreset{equation}{section}
\makeatother

\def\Acal{\mathcal A}
\def\qd{Q}

\def \eqref #1{(\ref{#1})}
\def \1{\mathbf 1}
\def \br{\begin{remark}}
\def\er{\end{remark}}

\def\vt\tilde{v}

\def\C{{\mathbb C}}
\def\Z{{\mathbb  Z}}
\def\R{{\mathbb R}}%%%%%%%%%%%%%%%%%%%%%%%%%%%%%%%%%%%%%%%%%%%%%%%%%%%%%%%%%%%%%%%%%%%%%%%%%%
\def\a{\alpha}
\def\g{\gamma}
\def\b{\beta}
\def\ka{\kappa}

\def\s{\sigma}
\def\p{\partial}
\def\Ccal{{\mathcal C}}
\def\Ch{{\widehat{{\mathcal C}}}}

\def\Mcal{{\mathcal M}}

\def\gt{{\tilde{\gamma}}}

\def\gh{\widehat{g}}

\def \Re{\mathrm {Re}\,}
\def \Im {\mathrm {Im}\,}
\def\f{\frac}
\def\la{\label}
\def\Scal{{\mathcal S}} % for Schwarzian derivative
\def\Rcal{{\mathcal R}}

\begin{document}

\title{Yang-Yang generating function and Bergman tau-function}

\vspace{0.2cm}
\begin{center}
\begin{huge}
%\fontfamily{cmss}
%\fontsize{17pt}{27pt}
%\selectfont
{WKB expansion of Yang-Yang generating function and Bergman tau-function}
\end{huge}\\
\bigskip
M. Bertola$^{\dagger\ddagger}$\footnote{Marco.Bertola@\{concordia.ca,sissa.it\}},  
D. Korotkin$^{\dagger}$ \footnote{Dmitry.Korotkin@concordia.ca},
\\
\bigskip
\begin{small}
$^{\dagger}$ {\it   Department of Mathematics and
Statistics, Concordia University\\ 1455 de Maisonneuve W., Montr\'eal, Qu\'ebec,
Canada H3G 1M8} \\
\smallskip
$^{\ddagger}$ {\it  SISSA/ISAS,  Area of Mathematics\\ via Bonomea 265, 34136 Trieste, Italy }\\
\end{small}
\vspace{0.5cm}
\end{center}

%
%\author{M.Bertola, D.Korotkin}
%\address{Department of Mathematics and
%Statistics, Concordia University\\ 1455 de Maisonneuve W., Montr\'eal, Qu\'ebec,
%Canada H3G 1M8}
%%\curraddr{}
%\email{Marco.Bertola@\{concordia.ca, sissa.it\}, Dmitry.Korotkin@concordia.ca}
%

%\baselineskip 14pt plus 1pt minus 1pt

%\subjclass[2010]

{\small 2010 Mathematics Subject Classification.  Primary 53D30, Secondary 34M45}

\begin{abstract}
We study the symplectic properties of the monodromy map of  second order equations on a Riemann surface whose   potential is meromorphic with double poles. We show that the  Poisson bracket defined in terms of  periods of meromorphic quadratic differential   implies the Goldman Poisson structure on 
the monodromy manifold.
These results are applied to the WKB analysis of the equation.
It is shown  that the leading term in the WKB expansion of the generating function of the monodromy symplectomorphism 
 (the ``Yang-Yang function" of Nekrasov, Rosly and Shatashvili \cite{NRS})
is determined by the Bergman tau-function on the moduli space of meromorphic quadratic differentials.

\end{abstract}

\tableofcontents
%\maketitle

\section{Introduction}

\rm

In this paper we study the symplectic properties and WKB expansion of the linear second order 
equation on a Riemann surface $\CC$ of genus $g$. This equation in a local coordinate $\xi$ can be written as
\be
\varphi_{\xi\xi}- u(\xi) \varphi =0\;.
\la{ped}
\ee
The equation (\ref{ped}) is invariant under the change of the local coordinate $\xi$ if the potential $u$ transforms as $1/2$ of a coordinate realization of a projective connection and the solution $\varphi$ transforms as the coordinate realization of a $-1/2$ differential \cite{HawSch}. The way to write the 
equation (\ref{ped}) in  coordinate-invariant form is to introduce the ``oper" $\partial^2- U$ where $U=u(\xi)(\d\xi)^2$,
and define
\be
\phi=\varphi(\xi)(\d\xi)^{-1/2}\;.
\ee
Then the equation (\ref{ped}) can be written in a coordinate-independent form as follows:
\be
(\partial^2- U)\phi=0
\la{oper}
\ee
where $U$ is $1/2$ of a projective connection on $\CC$. The operator $\partial^2$ acts on local sections of the line bundle $K^{-1/2}$  where $K$ is the canonical line bundle.
For detailed discussion of precise mathematical meaning of the equation (\ref{oper}) we refer to \cite{BeiDri}.

We shall parametrize the space of potentials of equation (\ref{oper}) by writing
$$U= -\f{1}{2} S_B+Q$$ 
where $Q$ is a meromorphic quadratic differential with simple zeros;
$S_B$
is the {\it Bergman projective connection}  which is a holomorphic projective connection \cite{Fay73} which  depends (locally and  holomorphically) only on the moduli of the Riemann surface $\CC$. The Bergman projective connection depends on the Torelli marking of the Riemann surface; it  is defined in terms of the canonical bimeromorphic differential $B(x,y)=\d_x\d_y\log E(x,y)$ (here $E(x,y)$ is the prime-form \cite{Fay73}) as follows:
\be
S_B(x) = \left(B(x,y)-\f{\d \xi(x) \d\xi(y)}{(\xi(x)-\xi(y))^2}\right)\Big|_{y=x}\;.
\la{SBint}
\ee

Then, writing $Q(\xi)=q(\xi) (\d\xi)^2$ and $S_B(\xi)=s_B(\xi) (\d\xi)^2$ we can represent the equation (\ref{ped}) in the form
\be
\varphi''+\left(\frac{1}{2}s_B-q\right)\varphi=0
\la{Sint1} 
\ee
while the coordinate-invariant equation (\ref{oper}) is written as 
\be
\partial^2\phi+\left(\frac{1}{2}S_B-Q\right)\phi=0\;.
\la{Sint} 
\ee

%Say that $\partial^2+\left(\frac{1}{2}S_0-Q$ is an "oper".

The differential $Q$  is assumed to have  $n$ second order 
poles $z_1,\dots,z_n$ on $\CC$ with fixed biresidues which we denote by $r_1^2,\dots,r_n^2$. Namely, 
\be
Q(x)\sim \left(\f{r_j^2}{\xi_j^2} + O(\xi_j^{-1})\right)(\d\xi_j)^2
\la{asQ}
\ee
as $x\to z_j$, where $\xi_j$ is a local coordinate near $z_j$.
%In (\ref{Sint}) $S_0$ is a holomorphic projective connection on $\CC$ which locally depends only on moduli of $\CC$,
%and $Q$ is a meromorphic quadratic differential on $\CC$ with biresidues $r_j^2$ at $z_j$'s and simple zeros. 
The space of pairs $(\CC,Q)$, such that all zeros of $Q$ are simple,  is denoted by  $\Qcal_{g,n}$. This space is foliated into 
leaves
  $\Qcal_{g,n}[{\bf r}]$ which correspond to fixed $r_j$'s.
The number of (simple) zeros of $Q$ is  $N=4g-4+2n$; we denote them by $x_j$, $j=1,\dots, N$.

%\footnote{Notice the minus sign in front of $Q$ in comparison with plus in \cite{BKN}}

%The coordinate invariance of the  equation (\ref{Sint})  implies  that  the solution $\phi$ locally transforms as $-1/2$-differential \cite{HawSch} under a coordinate change.

The equation  (\ref{Sint}) has Fuchsian singularities at poles of $Q$.
Assume that one of the solutions of (\ref{Sint}) has the asymptotics 
$$\phi_1=\xi_j^{\alpha_j}(1+O(\xi_j)) (\d\xi_j)^{-1/2}$$
 near $z_j$
where $\xi_j$ is a local coordinate near $z_j$ with $\alpha_j\not \in \Z/2$ (this assumption is needed to exclude the resonant case); then  the second solution
$\phi_2=\phi_1\int \phi_1^{-2}$
has the following asymptotics near $z_j$: 
$$\phi_2=\xi_j^{1-\alpha_j}(1+O(\xi_j))(\d\xi_j)^{-1/2}\;.$$

The singular part of the quadratic differential $Q$  at $z_j$ looks as follows:
\be
Q=\left(\f{\a_j(\a_j-1)}{\xi_j^2}+ O(\xi_j^{-1})\right) (d\xi_j)^2\;.
\la{singQ}
\ee
Thus, comparing with (\ref{asQ}), we have
\be
r_j^2=\a_j(\a_j-1)
\la{ra}
\ee
and
\be
\alpha_j=\f{1}{2}\pm r_j \sqrt{1+\f{1}{4r_j^2}}\;.
\la{alphar}
\ee
Here and below we assume that $r_j^2\not\in \R_-$ (which is necessary  for $Q$ to be free from  
saddle trajectories) and
choose the sign of $r_j$ such that
$$
\Re r_j <0\;.
$$
 
Let us introduce the monodromy group of equation (\ref{Sint}). 
 %the associate 
%Schwarzian equation
%for the ratio $f=\phi_1/\phi_2$ of two linearly independent solutions of the  equation (\ref{eq1}):
%\be
%\Scal(f,\xi)=S_0(\xi)-2Q(\xi)
%\la{SE}\ee
% where $\xi$ is an arbitrary local parameter on $\Ccal$ and $\Scal$ denotes the Schwarzian derivative:
% \be
 %\Scal(f,\xi)=\left(\f{f'}{f}\right)'-\f{1}{2}\left(\f{f'}{f}\right)^2\;.
% \la{Schwarzdef}
 %\ee
For each $Q\in \Qcal_{g,n}$ the ratio  $f=\phi_1/\phi_2$ transforms under analytic continuation
along a closed contour $\gamma$ as follows:
\be
f\to \f{a_{\gamma} f+ c_{\gamma}}{b_{\gamma} f+ d_{\gamma}}\;\hskip0.7cm {\rm where}\hskip0.7cm
M_\gamma=\left(\ba{cc} a_\gamma & b_\gamma \\ c_\gamma & d_\gamma \ea\right)^{-1}
\la{fral}
\ee
which determines a
$PSL(2,\C)$ monodromy representation of the fundamental group 
$\pi_1(\C\setminus\{z_i\}_{i=1}^n,\,x_0)$ with $x_0\in \CC$ being a base point.

 Denote the standard generators of the fundamental group by
$\a_1,\b_1,\dots,\a_g,\b_g,\ka_1,\dots,\ka_n$; these generators satisfy the relation 
\be
\ka_1\dots\ka_n\prod_{i=g}^1 \a_i\b_i\a_i^{-1}\b_i^{-1}=id
\la{relfun}\ee
where the loop $\ka_j$ goes around $z_j$ counter-clockwise.
The $PSL(2)$ monodromy matrices satisfy the same relation
\be
M_{\ka_1}\dots M_{\ka_n} \prod_{i=1}^g M_{\a_i} M_{\b_i} M_{\a_i}^{-1} M_{\b_i}^{-1} =I\;.
\la{relmon}
\ee

 The diagonal form of the matrix $M_{\ka_j}$ is given by
 \be
 \Lambda_j=\left(\ba{cc} m_j & 0\\
 0 & m_j^{-1} \ea\right)
 \la{Lambdaj}
 \ee
where 
\be
m_j^2= e^{4\pi i \a_j}\;.
\la{ma}
\ee
%are defined up to a sign (as well as all the matrices $M_\gamma$). 

%related to biresidues $r_j^2$ of the differential $Q$ via
%\be
 %r_j^2=\f{\log m_j}{2\pi i}\left(\f{\log m_j}{2\pi i}-1\right)
 %\la{resmon}
%\ee
%\red{sign of $m_j$??}

The key analytical object associated to the equation (\ref{Sint}) is  the canonical cover $\Ch$ defined by $v^2=Q$.
%\la{ccint}
%\ee
The cover $\Ch$ is equipped with the natural holomorphic involution $\mu$ 
and the canonical projection $\pi:\Ch\to \CC$.
In this paper we assume that all zeros of $Q$ are simple.
Then the genus of $\Ch$ equals to $\gh=g+g_-$ with $g_-=3g-3+n$. 

Denote two points of $\Ch$ projecting to the pole $z_j$ by $z_j^{(1)}$ and $z_j^{(2)}$. The enumeration of these points is chosen such that the residue of $v$ at $z_j^{(1)}$ equals $r_j$ and 
the residue of $v$ at $z_j^{(2)}$ equals $-r_j$.

The homology group 
$H_1(\Ch\setminus\{z_j^{(1)},z_j^{(2)}\}_{j=1}^n)$  splits into the direct sum $H_+\oplus H_-$ of two eigenspaces of the operator $\mu_*$.
Then ${\rm dim }\; H_-= 6g-6+3n=2g_- +n$; one can choose generators of $H_-$ as follows:
\be
\{a_i^-,b_i^-\}_{i=1}^{g_-},\;\;\{t_l^-\}_{l=1}^n\;.
\la{genin}\ee
where the cycles $t_j^-$ is $1/2$  of the difference of two small counter-clockwise circles around 
$z_j^{(1)}$ and $z_j^{(2)}$.
The intersection indices between these cycles are given by $a_j^-\circ b_k^-=\delta_{jk}/2$; cycles $t_j^-$ have trivial intersection indices with all cycles. 

Trivially, $\int_{t_j^-}v = r_j$.
The integrals of $v$ over the remaining cycles 
$$ A_j =\int_{a_i^-}v\;, \hskip0.7cm  B_j =\int_{b_i^-}v$$
for $j=1,\dots g_-$  can be used as local {\it period} (or {\it homological}) coordinates on $\Qcal_{g,n}[{\bf r}]$.
We introduce the homological Poisson bracket on $\Qcal_{g,n}$ as follows \be
\left\{\int_{s_1}v\,,\, \int_{s_2} v\right\}_{hom} = s_2\circ s_1
\la{homPB}
\ee
for any $s_1, s_2\in H_-$ where $\circ$ is the intersection index in $H_-$.
The space $\Qcal_{g,n}[{\bf r}]$ is a symplectic leaf of the bracket (\ref{homPB}).
The following {\it homological} symplectic form on  $\Qcal_{g,n}[{\bf r}]$ is the inverse of (\ref{homPB}):
\be
\Omega_{hom}=2\sum_{j=1}^{g_-} \d B_j\wedge \d A_j\;.
\la{homin}
\ee
It was proved in \cite{BKN,GT} that in the case of $n=0$ (when $Q$ is holomorphic and can be identified with the cotangent vector to the moduli space ${\mathcal M}_g$ of Riemann surfaces of genus $g$) the homological 
symplectic form coincides with the canonical symplectic form on $T^* {\mathcal M}_g$. This result was generalized in \cite{Kor} to the case when $Q$ has $n$ simple poles; in that case the homological form 
coincides with the canonical symplectic form on $T^* {\mathcal M}_{g,n}$. We are not aware of a similar result in the present case, when $Q$ has $n$ poles of second order.

Let us choose the  symplectic potential $\theta_{hom}$ (such that $\d\theta_{hom}=\Omega_{hom}$) as follows:
\be
\theta_{hom}=\sum_{j=1}^{g_-}  B_j \d A_j - A_j \d B_j \;.
\la{thin}
\ee

Denote now the $PSL(2)$ character variety of $n$-punctured Riemann surface of genus $g$ by $CV_{g,n}$.
Denote by $CV_{g,n}[{\bf m}]$ the leaf of $CV_{g,n}$ such that the eigenvalues of the   monodromy matrix
around $z_j$ are given by $m_j$ and $m_j^{-1}$. % with $\a_j$ being one of two solutions of (\ref{ra}).
The natural symplectic form $\Omega_G$  on $CV_{g,n}[{\bf m}]$ is given by the inverse of the 
 $PSL(2)$ Goldman bracket.

 An important technical result is  (see App. \ref{app2}):
 \begin{proposition}
 The symplectic form  $\Omega_G$ inverting the Goldman's bracket on a symplectic leaf
  $CV_{g,n}[{\bf m}]$
 can be written in the canonical form
 \be
 \Omega_G= 2\sum_{j=1}^{g_-}  \d \FG_{a_j^-}\wedge  \d \FG_{b_j^-}
 \la{Gint}
 \ee
where the Darboux coordinates $\FG_{a_j^-}$ and $\FG_{b_j^-}$    are appropriate linear combinations of (logarithms of) Thurston's shear coordinates;
%(the $PSL(2)$ case of more general Fock-Goncharov coordinates); 
we call them the 
{\it homological shear coordinates}  on  $CV_{g,n}[{\bf m}]$, see (\ref{abl}), (\ref{shearperiods})  for their detailed definition.
\end{proposition}

The  symplectic potential $\theta_G$ (such that $\d\theta_G=-\Omega_G$) can be chosen as follows:
\be
\theta_G=\sum_{j=1}^{g_-}  \FG_{b_j^-} \d \FG_{a_j^-} -  \FG_{a_j^-} \d \FG_{d_j^-}\;.
\la{potGin}
\ee

%The monodromy matrices of (\ref{Sint}) are defined to be the $PSL(2,\C)$ monodromies of the corresponding Schwarzian equation for $f=\phi_1/\phi_2$
%\be
%\Scal(f,\xi)=S_B(\xi)-2 Q(\xi)
%\la{SEin}
%\ee
%where $\Scal$ denotes the Schwarzian derivative.

The first main result of this paper is the following
 \begin{theorem}
The monodromy map
$$       \Fcal: \Qcal_{g,n} \to CV_{g,n}      $$
 for equation (\ref{Sint}) is Poisson. The Poisson structure on $\Qcal_{g,n}$ is given by 
 (\ref{homPB}) while the Poisson structure on $CV_{g,n} $ is defined by minus the Goldman's bracket (\ref{Goldint}).
\end{theorem}

It follows from the theorem that the restriction of $\Fcal$ to the corresponding symplectic leaf
$$\Fcal: \Qcal_{g,n}[{\bf r}] \to CV_{g,n}[{\bf m}]$$
(where $m_j$ is expressed in terms of $r_j$ via (\ref{alphar}), (\ref{ma}))
is a symplectomorphism, and, therefore,   the 1-form $\Fcal^* \theta_G- \theta_{hom}$ is closed.

\begin{definition}
The generating function of the symplectomorphism $\Fcal$ is  defined by
\be
\d\Gcal= \Fcal^* \theta_G- \theta_{hom}\;.
\la{GG}
\ee
\end{definition}

The function $\Gcal$ depends on the choice of canonical basis of cycles on $\CC$ (the Torelli marking)
used to define the Bergman projective connection $S_B$,  on the choice of triangulation $\Sigma$ of $\CC$ 
(see Appendix \ref{app2} for details)
and a basis in $H_-$ used to define the homological shear coordinates $\FG_{a_j^-}$, 
$\FG_{b_j^-}$ on $CV_{g,n}[{\bf m}]$.

Our second main result is the computation of the leading term of the WKB expansion of the function $\Gcal$.
Namely, we introduce a small real positive parameter $\hbar$ in the equation (\ref{Sint})  as follows

\be
\partial^2 \phi +\left(\frac{1}{2}S_B-\f{1}{\hbar^2}Q\right)\phi=0\;.
\la{Sinth} 
\ee

%Then the WKB expansion produces a family of meromorphic differentials on $\Ch$ which are skew-symmetric under the involution $\mu$ and have singularities only at the poles and zeros of $Q$.
%\red{more details!}
%The first two of these differentials are $v_{-1}$ and $v_1$; one can choose $v_{-1}=v$; then $v_1$ is given by the following formula:
%\be
%v_1=\f{S_B-S_v}{2v}
%\la{v1}
%\ee
%where $S_v=\Scal(\int^x v, \xi(x))(\d\xi(x))^2$ is the meromorphic projective connection on $\CC$ defined by 
%the differential $v$.
The WKB ansatz for the solution of this equation is
\be
\phi=v^{-1/2} \exp\left\{\int_{x_0}^x (\e^{-1} s_{-1}+s_0+\e s_1+\dots)v\right\}
\la{WKBint}
\ee
where $v_j=s_j v$ are meromorphic differentials on $\Ch$. Of particular importance is the sum of odd-numbered differentials:
\be
v_{odd}=s_{odd} v = \sum_{k=-1}^\infty s_{2k+1} v \;;
\la{soddint}
\ee
the integrals of the differentials $v_{2k+1}=s_{2k+1} v$ over cycles of $H_-$ are called {\it Voros symbols}.

In Section \ref{WKBsec} we find the   WKB expansion of the  %logarithmic 
homological shear coordinates. Assume that the differential $Q$ is a GMN (Gaiotto-Moore-Nietzke) differential i.e. 
it is free from saddle connections (none of horizontal trajectories of $Q$ connect two of its zeros). 
Any horizontal trajectory  emanating from a zero of $Q$ ends at a pole; such trajectories are called {\it critical}. Three critical trajectories meet at each zero (recall that all zeros of $Q$ are assumed to be simple).

The set of critical trajectories defines two natural graphs embedded in $\CC$. The graph 
$\Sigma_Q$  is a triangulation of $\CC$ whose vertices are poles $z_j$; each triangle of $\Sigma_Q$ contains one zero $x_k$ such that the critical trajectories emanating from $x_k$ go towards the vertices of the corresponding triangle.   The  dual tri-valent graph $\Sigma_Q^*$  has $n$ faces (and each face contains one pole of $Q$) and $4g-4+2n$ vertices at zeros $x_j$.

%We prove that the graph $\Sigma_Q^*$ can be used to make an explicit construction of the canonical covering $v^2=Q$ which is summarized in the following technical statement:
\begin{proposition}
Assume that all edges of the graph $\Sigma_Q^*$ are branch cuts and glue two copies of $\CC$ together along them. Then the resulting covering is holomorphically equivalent to the canonical covering 
$v^2=Q$.
\end{proposition}

This proposition is  non-trivial because  one can get many inequivalent branch coverings
by choosing different sets of edges of $\Sigma_Q^*$ as branch cuts (such that an odd number of branch cuts - 1 or 3 meets at each vertex $x_k$). Some of those coverings are equivalent to each other: in every equivalence class there are $2^{n-1}$ 
different configurations of branch cuts.
%, but the total number of coverings one can get via this construction is much higher.

The graph $\Sigma_Q$ is  used in App. \ref{monre} to define the set of Thurston's (logarithmic) shear coordinates and their 
homological analogs. % (Thurston's shear coordinates are $SL(2)$ special case of general Fock-Goncharov coordinates).  
The original shear coordinates are naturally assigned to each edge of the graph $\Sigma_Q$
(or $\Sigma_Q^*$); we extend them to any cycle $\ell\in H_-$ by linearity and denote the corresponding logarithmic shear coordinate, which we call homological, by $\FG_\ell$.

\begin{proposition} (Prop. \ref{asFG})
Let $Q$ be a GMN differential.
Denote by $\FG_\ell$ the (logarithmic) homological shear coordinate corresponding to a cycle $\ell\in H_-$.
Then we have the following asymptotic expansion (in Poincar\'e sense) of $\FG_\ell$  as $\hbar\to 0^+$: 
\be
\FG_\ell\sim  \int_\ell v_{odd} 
%v +\hbar \int_{l} v_1 + \hbar^3 \int_{l} v_3+\dots 
\la{zetaas}
\ee
where $v_{odd}=s_{odd} v$ is given by (\ref{soddint}) as the formal series in $\hbar$.
\la{th1in}
\end{proposition}
The asymptotic expansion (\ref{zetaas}) is similar to the asymptotic expansion of  shear coordinates in terms of Voros symbols proved in \cite{AlBrid,Alleg}. However, it is not quite the same statement since in \cite{Alleg} the reference projective connection is supposed to have second order poles at $z_j$'s while the Bergman projective connection which is used as  reference point in this paper is holomorphic on $\CC$.

For any two differentials $v,w$ on $\Ch$ with poles (possibly with residues) at $z_1,\dots,z_n$ 
and a chosen set of generators (\ref{genin}) in $H_-$
 we introduce the pairing
 $$
\langle v, w\rangle =\sum_{j=1}^{g_-}\left[\int_{b_j^-} v\;\int_{a_j^-} w-
\int_{a_j^-} v\;\int_{b_j^-} w\right]\;.
%\la{pairingint}
$$

Now substitution of the expansion (\ref{zetaas}) into the definition of the generating function $\Gcal$ (\ref{GG})
leads to the following theorem (Th.\ref{mainth}):
\begin{theorem}
The leading term of the asymptotic expansion ($\hbar\to 0^+$) of  the generating function $\Gcal$ (\ref{GG})
%of the monodromy symplectomorphism of equation (\ref{eqhbar1}) 
has the following form
\be
\Gcal= -12\pi i\, \log \tau_B +2 \langle v, v_1\rangle + O(\hbar^2 )
\la{asYYint}
\ee
where $\tau_B$ is the Bergman tau-function (\ref{taupint}) on the moduli space of quadratic differentials 
with second order poles (see more detail in Sec.\ref{tausec});
the differential $v_1$ is given by 
$$
v_1=s_1 v=\f{\Scal_B-\Scal_v}{2v}
$$
where 
\be
S_v=\Scal\left(\int^x v\,,\, \xi(x)\right) (\d\xi(x))^2\;
\la{Sv}\ee
($\Scal$ is the Schwarzian derivative  and $\xi(x)$ a local coordinate on $\CC$). 
\end{theorem}

The Bergman tau-function can be interpreted as determinant of a $\bar{\partial}$ operator in the spirit of
Quillen \cite{Quillen} acting on functions on $\CC$ (more precisely, $\tau_B$ is a section of the
determinant line bundle of the Hodge vector bundle over  the moduli space of quadratic differentials with $n$ second order poles with fixed biresidues over Riemann surfaces of genus $g$; it enters also the formula of holomorphic factorization of determinant of Laplace operator \cite{JDG}). In physics terms, $\tau_B$ can be interpreted as chiral partition function of free bosons on a Riemann surface (see \cite{Knizhnik,HP} and references therein).
Therefore, the formula (\ref{asYYint}) can be also interpreted as follows:
\be
\Gcal= -12\pi i\, \log {\rm det}\bar{\partial} +2 \langle v, v_1\rangle + O(\hbar^2 )\;.
\la{ttint}
\ee

This paper leads to many interesting questions (see Sec.\ref{openpro}). Here
we only mention one of them: what is the role of the higher order terms in the WKB expansion of $\Gcal$?
The most naive conjecture would be to expect that these terms coincide with higher genus free energies 
in  (the appropriate version of) the framework of topological recursion \cite{EO}.
 
\section{Quadratic differentials with second order poles}

\subsection{Geometry of canonical cover}

%Assume that the singular part  of $Q$ near the pole $z_j$ is given by (\ref{asQ}) 
%($\xi_j$ is a local coordinate near $z_j$):
%\be
%Q(\xi_j)=\left(\frac{r_j^2}{\xi_j^2} + \dots\right)(\d \xi_j)^2
%\la{sinQ}
%\ee
%and $r_j^2\not \in \R_-$.

Denote  by $\Qcal_{g,n}$ the moduli space of  meromorphic quadratic differentials on a Riemann surface of genus $g$ with $n$ double poles and $4g-4+2n$
simple zeros. 
We assume that the singular part  of $Q$ near the pole $z_j$ is given by (\ref{asQ}); for the purposes of the WKB expansion  we shall need to assume that the biresidues $r_j^2\not\in \R_-$.
For any differential $Q\in \Qcal_{g,n}$ we define the canonical covering $\Ch$ as the locus  in $T^* \Ccal$ given   by
\be
v^2=Q\;.
\la{cancov}
\ee
The two-sheeted covering $\pi:\Ch\to\Ccal$ is branched at  zeros of $Q$; thus the total number of branch points is 
$4g-4+2n$ and the  genus of $\Ch$ equals $\gh=4g-3+n$.  Recall that  the natural involution on $\Ch$ is denoted by $\mu$ and the projection 
of $\Ch$ to $\Ccal$ by $\pi$.
The Abelian differential $v$ on $\Ch$ is  of third kind  with   double  zeros   at the  branch points $\{x_i\}_{i=1}^{4g-4+2n}$; it is skew-symmetric under the involution $\mu$. It
has simple poles at $2n$ points which 
 we denote by $z_i^{(1)}$ and $z_i^{(2)}$ with  residues $r_j$ and $-r_j$, respectively. We shall fix the sign of $r_j$ by the assumption that $\Re r_j <0$.

Consider the homology group 
\be
H_1(\Ch\setminus\{z_i^{(1)},z_i^{(2)}\}_{i=1}^n,\R)\;.
\la{homg}\ee
 The generators of this group can be chosen as follows (this is the generalization of the basis used in \cite{Fay73,contemp,BKN} to the case $n\neq 0$):
\be
\{a_j,  a_j^\mu, \at_k,\,b_j,  b_j^\mu\,,\bt_k, t_l, t_l^\mu\}\,,\quad j=1,\dots,g,\;\;k=1,\dots,2g-3\;,\;\; l=1,\dots, n
\label{mainbasis}
\ee
where $\{a_i,b_i,a_i^\mu,b_i^\mu\}_{i=1}^g$ is a lift   to $\Ch$ of the canonical basis of cycles $\{a_i,b_i\}_{i=1}^g$  on $\CC$,
such that 
\be 
\mu_* a_j = a_j^\mu,\quad\mu_* b_j = b_j^\mu,
\quad\mu_* \at_k + \at_k= \mu_* \bt_k + \bt_k  = 0\;;
\ee
$\{t_l, t_l^\mu\}$ is a lift to $\Ch$ of the  small positively-oriented circle $t_l$ around $z_l$ on $\CC$.  We define  
$t_l$ to be  a small positively oriented loop  around $z_j^{(1)}$; then $t_l^\mu$ is a small loop around $z_j^{(2)}$.
In the group (\ref{homg}) one has the relation
\be
\sum_{l=1}^n (t_l+t_l^\mu) =0\;.
\la{relhom}
\ee

\begin{figure}
\begin{center}
\includegraphics[width=0.6\textwidth]{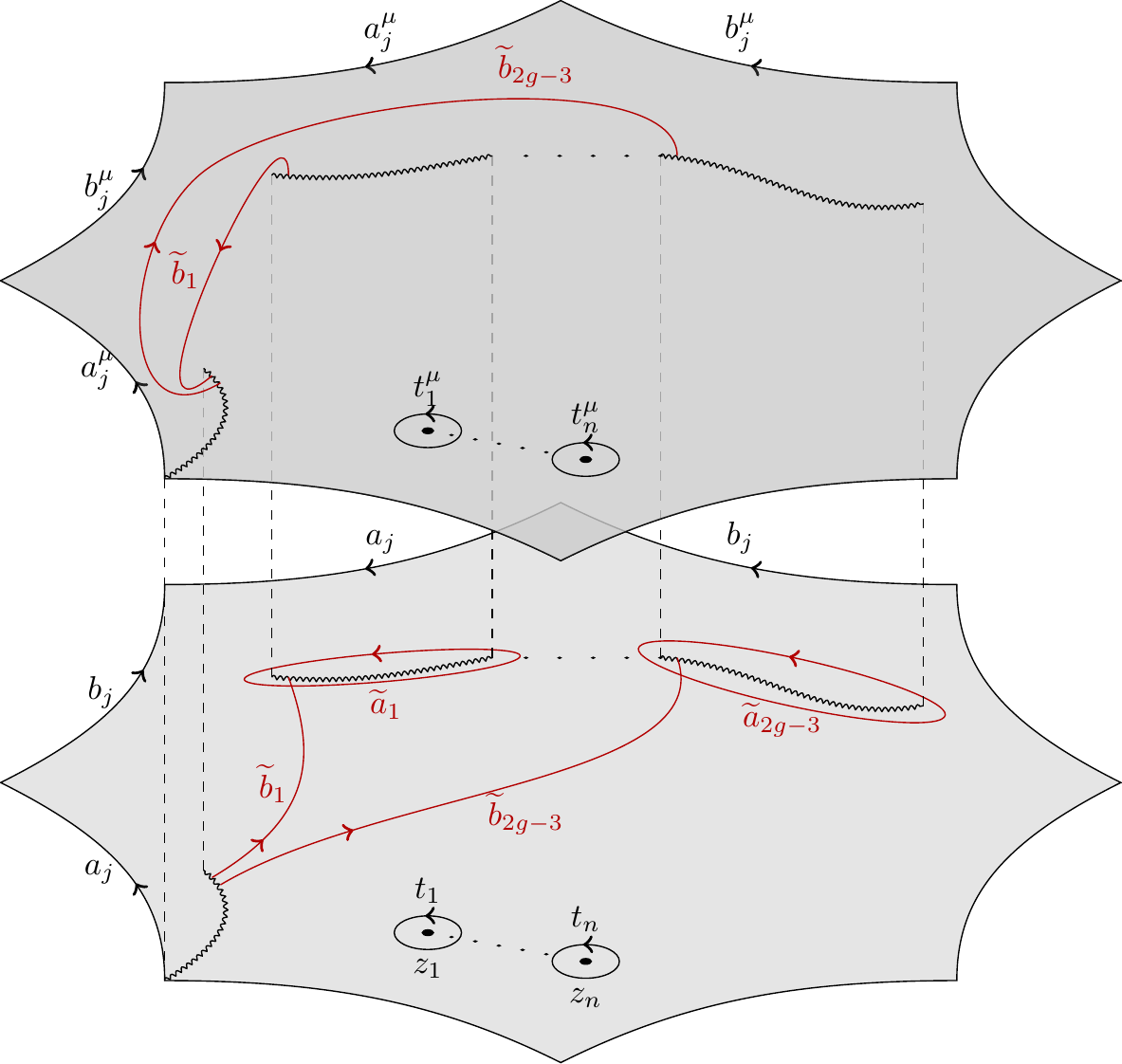}
\end{center}
\caption{Generators of  $H_1(\Ch\setminus\{z_j^{(1)},z_j^{(2)})$.}
\label{figBasis}
\end{figure}

Let us  decompose the homology group (\ref{homg}) into symmetric and skew-symmetric parts under the action of the  involution $\mu_*$:
$$H_1(\Ch\setminus\{z_i^{(1)},z_i^{(2)}\}_{i=1}^n)=H_+\oplus H_-$$
 where 
 $\dim H_+=2g+n-1$ and $\dim H_-=6g-6+3n\;$. 
 %The holomorphic part of cohomology group $H^{(1,0)}(\Ch,\R)$ is similarly decomposed as $H^+\oplus H^-$ where $\dim H^+=g$ and $\dim H^-=3g-3+n$.

The classes in (\ref{homg}) given by 
\be
a_j^+ = \f{1}{2}(a_j+ a_j^\mu)\;,\quad
b_j^+ = \f{1}{2}( b_j+ b_j^\mu)\;,\quad j=1,\dots,g,\hskip0.7cm
t_l^+= \f{1}{2}(  t_l+t_l^\mu)\;,\quad l=1,\dots,n
\label{abp}
\ee
are generators of  $H_+$. They have  intersection indices 
$$
a_j^+\circ b^+_k=\f{1}{2}\delta_{jk}
$$
(while $t_l^+$ have vanishing intersection with all cycles).

Introduce a set of free generators of $H_-$ denoted by 
\be
\{a_i^-,b_i^-\}_{i=1}^{3g-3+n}\;,\;\;\;\{t_l^-\}_{l=1}^n
\la{homba}\ee
and defined as follows:
\be
a_l^- =  \f{1}{2}(a_{l}- a_{l}^\mu) ,\hskip0.5cm
b_l^- = \f{1}{2}( b_{l}- b_{l}^\mu),\quad l=1,\dots,g\;,
\la{abm1}
\ee
\be
a^-_{l}=\f{1}{\sqrt{2}}\at_{l-g},\;\;\; b^-_{l}=\f{1}{\sqrt{2}}\bt_{l-g}\;,\hskip0.5cm\, l=g+1,\dots,3g-3+n\;;
\label{abm}
\ee
the generators $t_l^-$ are given by
\be
t_l^-=\f{1}{2}( t_l-t_l^\mu)\;.
 \la{tjm}
 \ee
 
These classes have the following  intersection indices:
\be
a_i^-\circ b_j^-=\f{\delta_{ij}}{2}\;,\hskip0.7cm
%\la{intind}
\ee
and all other intersection indices (in particular, the intersection indices of $t_i^-$ with all  generators) equal to $0$. 

 The rank of the intersection pairing on $H_-$ equals $6g-6+2n$.

\subsection{Period coordinates. Homological  Poisson structure and symplectic form}

%Denote by $\Qcal_{g,n}$ the moduli space of quadratic differentials with $n$ second order poles and 
%all simple zeros on Riemann surfaces of genus $g$.

The dimension of the space $H_-$ coincides with the dimension of the moduli space $\Qcal_{g,n}$.
This allows to define the set of {\it period} (or {\it homological}) local coordinates on $\Qcal_{g,n}$ as follows:

\be
A_i=\int_{a_i^-} v\;\hskip0.7cm B_i=\int_{b_i^-} v\;, \hskip0.7cm
2\pi i r_k=\int_{t_k^-} v
\la{homco}
\ee
for $i=1,\dots 3g-3+n$ and $k=1,\dots,n$.

We denote by $\Qcal_{g,n}[{\bf r}]$ the stratum of $\Qcal_{g,n}$ corresponding to fixed values of $r_1,\dots,r_n$. The periods $\{A_i,B_i\}_{i=1}^{3g-3+n}$ can be used as local {\it homological} coordinates on $\Qcal_{g,n}[{\bf r}]$.

We define the homological Poisson structure on $\Qcal_{g,n}$ in terms of periods (\ref{homco}):
\begin{definition}
For two cycles $\ell_1,\ell_2\in H_-$ the homological Poisson structure between corresponding periods of $v$ is given by
the intersection index of the  cycles:
\be
\left\{\int_{\ell_1}v\;,\,\int_{\ell_2} v\right\}=\ell_2\circ \ell_1\;.
\la{PBhom}
\ee
\end{definition}

The Poisson structure (\ref{PBhom}) is degenerate; the Casimirs are given by $r_1,\dots,r_n$. The leaves
$\Qcal_{g,n}[{\bf r}]$ are symplectic for the bracket (\ref{PBhom}). 
%The symplectic structure on them is obviously given by
\begin{proposition}
The homological symplectic form on $\Qcal_{g,n}[{\bf r}]$ is defined by
\be
\Omega_{hom}=\;\;\; 2\sum_{i=1}^{g_-} \d B_i\wedge \d A_i\;.
\la{defsym}\ee
\end{proposition}

\begin{remark}\rm
The coordinates used in  \cite{BKN}  in the case $n=0$ were equal to  $i A_j$ and $i B_j$
due to a different sign in front of $Q$ in (\ref{Sint}), which implies the change of order in the form (\ref{defsym}) in comparison with \cite{BKN}.
\end{remark}

\begin{remark}\rm
The real slice  $\Qcal_{g,n}^{\R}[{\bf r}]$    of $\Qcal_{g,n}[{\bf r}]$, where all     
periods of $v$ in $H_-$ are real (thus in particular all $r_j$ are imaginary)
coincides with the largest stratum of the combinatorial model based on Strebel differentials \cite{Kon,BK1}. The symplectic form (\ref{defsym}) restricted to $\Qcal_{g,n}^{\R}[{\bf r}]$ coincides with the symplectic form used in \cite{Kon} to introduce the orientation on the largest stratum of the combinatorial model. This restriction of the  form $\omega$  (see
 \cite{Kon}) equals to $\sum_{j=1}^n p_j^2 \psi_j$ where $\psi_j$ are tautological $\psi$-classes and $p_j=2\pi i r_j$ \cite{BK1}.
\end{remark}

% If $S_0$ is a Bergman projective connection $S_B^\a$ then $\Theta={\d}\log \tau_B$ where $\tau_B$ is the Bergman tau-function on the space of meromorphic quadratic differentials wih second order poles \cite{KalKor,BK1,KorRev}.
 
\section{Monodromy map }
%\section{Poisson brackets on monodromy manifold}

\subsection{Poisson brackets for potential and variational formulas on $\Qcal_{g,n}[{\bf r}]$}

%Let us now choose in (\ref{Sint}) $S_0$ to be the Bergman projective connection $S_B$ corresponding to some choice of the canonical basis of cycles. Then  (\ref{Sint}) takes the form
%\be
%\phi''+\left(\frac{1}{2}S_B-Q\right)\phi=0\;.
%\la{eq1} 
%\ee
 Denote two linearly independent solutions of (\ref{Sint}) by $\phi_1$ and $\phi_2$ and introduce functions $\varphi_{1,2}=\phi_{1,2}\sqrt{v}$.% where $\phi_{1,2}$ are two linearly 
%independent solutions of   (\ref{Sint}). 
 Denote by $\Psi$ the Wronskian matrix of 
$\varphi_1$ and $\varphi_2$. The matrix $\Psi$ satisfies the first order matrix equation
\be
\d\Psi= \left(\ba{cc} 0 & v \\
uv & 0  \ea \right)\Psi\;,
\la{lsi}\ee
 where the {\it potential} (which is the   meromorphic function   on $\CC$) is given by 
\be
u= -\f{S_B-S_v}{2Q}+1
\la{potfor}\ee
and $S_v(\xi)=\Scal(z(\xi),\xi)$ is the Schwarzian derivative  of the  coordinate $z(x)=\int_{x_1}^x v$ with respect to a local coordinate $\xi$. The coefficients $v$ and $uv$ in (\ref{lsi}) are  meromorphic 
differentials on $\Ch$, which are skew-symmetric under the involution $\mu$.

The homological  symplectic structure (\ref{defsym}) on the space $\Qcal_{g,n}[{\bf r}]$
induces the Poisson structure on the space of coefficients $u$ of the equation (\ref{lsi}).  The Poisson bracket between $u(z)$ and $u(\zeta)$ (for constant $z$ and $\zeta$) can be computed using the definition 

\be
\{u(x),u(y)\}=\f{1}{2}\sum_{i=1}^{3g-3+n} \left(\frac{\p u(x)}{\p B_i} \frac{\p u(y)}{\p A_i}-\frac{\p u(x)}{\p A_i} \frac{\p u(y)}{\p B_i}
\right)
\la{uudef}
\ee
(where it is assumed that the flat coordinates $z(x)$ and $z(y)$ remain constant)
and variational formula (\ref{varQ}) below. 
This  gives, in analogy to the proof of Proposition 4.4 of \cite{BKN}:
\be
\f{4\pi i}{3}\{u(x), u(y)\}
=\L_z\left[\int^x h(z,\zeta)dz\right]-
\L_\zeta\left[\int^y h(z,\zeta)d\zeta\right]
\la{Poissonu}
\ee
where $z=z(x)$, $\zeta=z(y)$, $h(z,\zeta)=\frac{B^2(x,y)}{Q(x)Q(y)}$ and 
 $$
 \L_z=\f{1}{2}\p_z^3-2u(z)\p_z-u_z(z)
 $$
 is called the  ``Lenard's operator'' in the theory of the Korteveg de Vries equation. 
 %In the definition of the bracket (\ref{Poissonu})
%it is assumed that the arguments $z$ and $\zeta$ of $u$ are independent of moduli.

The proof of (\ref{Poissonu}) is identical to that in the case of holomorphic potentials (see Prop.4.4 of \cite{BKN}).
It uses variational formulas for the potential $u$ (\ref{potfor}) on the space $\Qcal_{g,n}[{\bf r}]$   which take the form  (see (3.35) of \cite{BKN} and Section 11  of \cite{KorRev}):
\be
\label{varQ}
\f{\p u(x)}{\p A_j}\Big|_{z(x)=const}=\f{3}{2\pi i}\int_{b_j^-} \f{B^2(x,y)}{v^2(x)v(y)}\;,\hskip0.8cm
\f{\p u(x)}{\p B_j}\Big|_{z(x)=const}=-\f{3}{2\pi i}\int_{b_j^-} \f{B^2(x,y)}{v^2(x)v(y)}\;.
\ee
Substituting  these formulas into (\ref{uudef}), using  the Riemann bilinear identities and computing the resulting contour integral by residues leads  to (\ref{Poissonu}).

\subsection{Monodromy symplectomorphism  and its generating  function}

%The ratio $f=\phi_1/\phi_2$ of two linearly independent solutions of the  equation (\ref{Sint})
 %solves the Schwarzian equation
%\be
%\Scal(f,\xi)=S_B(\xi)-2Q(\xi)\;,
%\la{SE}\ee
 %where $\xi$ is an arbitrary local parameter on $\Ccal$ and $\Scal$ denotes Schwarzian derivative.
%For each $Q\in \Qcal_{g,n}$ the Schwarzian equation determines a $PSL(2,\C)$ monodromy representation of the fundamental group 
%$\pi_1(\C\setminus\{z_i\}_{i=1}^n,\,x_0)$ (which turns out to be liftable to an $SL(2,\C)$ representation for the case $n=0$  \cite{GaKaMa}) for a basepoint $x_0\in\CC$.
%Consider  the $PSL(2)$ character variety $CV_{g,n}$ 
%which contains equivalence classes of monodromy representations (moduli simultaneous conjugations of all monodromies by the same matrix).

Consider the monodromy map
\be
\Fcal: \;\Qcal_{g,n}\to CV_{g,n}
\la{monmap}
\ee
for the equation (\ref{Sint}) defined by (\ref{fral}).
%is well-defined since the point in the character variety does not depend on the initial point $x_0$.
  
  %Traces of monodromy matrices along loops $\gamma\in \pi_1(\CC)$ can be used as local coordinates on
 % $CV_{g,n}$.
The Goldman bracket on $CV_{g,n}$ is defined as follows \cite{Gold84}.  For  two arbitrary loops $\g$ and $\gt$
\be
\{{\rm tr}M_\g,\;{\rm tr}M_\gt\}_G=\f{1}{2}\sum_{p\in \g\circ\gt}\nu(p)\left( {\rm tr}M_{\g_p \gt}- {\rm tr}M_{\g_p \gt^{-1}}\right)
\la{Goldint}
\ee
where the monodromy matrices $M_\g,M_\gt\in PSL(2,\C)$;  $\g_p \gt$ and $\g_p \gt^{-1}$ are  paths obtained by  
resolving the intersection point $p$ in two different ways (see \cite{Gold84}); $\nu(p)=\pm 1$ is the contribution of the point
$p$ to the intersection index of $\gamma$ and $\tilde{\gamma}$. 
%\footnote{\blue{The sign in the definition of Goldman's bracket we choose here in non-canonical; we choose this definition since it simplifies several computations below}}

The following theorem was proved in \cite{BKN} for holomorphic $Q$ and in \cite{Kor} for potentials with simple poles. The proof for the present case of second order poles is parallel (in \cite{BKN, Kor} we were working with the lift of the              $PSL(2)$ monodromy representation to $SL(2)$; in this paper we deal with $PSL(2)$ representation for technical simplicity).

\begin{theorem}
 The monodromy map $\Fcal$ (\ref{monmap})  of equation (\ref{Sint}) is Poisson. Namely, the homological Poisson bracket (\ref{PBhom}) implies minus the the  Goldman bracket (\ref{Goldint})      between traces of monodromy matrices of the
equation (\ref{lsi}).
\end{theorem}

Therefore, the monodromy map is a symplectomorphism between the symplectic leaves. Namely, denote by 
$CV_{g,n}[{\bf m}]$ the symplectic leaf of the character variety $CV_{g,n}$ corresponding to fixed eigenvalues 
$m_j$. We denote by
$\Omega_G$ the symplectic form on $CV_{g,n}[{\bf m}]$ which is the inverse of Goldman Poisson structure. 

The  monodromy map sends the leaf  $\Qcal_{g,n}[{\bf r}]$ to the leaf $CV_{g,n}[{\bf m}]$ with
$r_j$ and $m_j$ related by
\be
m_j^2={\rm exp}\left[\pm 4\pi i \sqrt{r_j^2+\f{1}{4}}\right]\;.
%r_j^2=\f{\log m_j}{2\pi i}\left(\f{\log m_j}{2\pi i}-1\right)\;.
\la{rm}
\ee
(the change of the sign in the r.h.s. corresponds to the interchange of $m_j$ and $m_j^{-1}$ i.e. the change of the ordering of the eigenvalues of $M_j$).

\begin{corollary} 

The homological symplectic form $\Omega_{hom}$ (\ref{defsym}) on $\Qcal_{g,n}[{\bf r}]$ is related to Goldman symplectic form $\Omega_G$ on $CV_{g,n}[{\bf m}]$ as follows:
\be
\Omega_{hom}= -\Fcal^*\Omega_G\;.
\la{pullback}
\ee
\end{corollary}

Since the monodromy map is a symplectomorphism between symplectic leaves in $\Qcal_{g,n}$ and
$CV_{g,n}$ one can define the generating function of this map (the ``Yang-Yang" function of 
\cite{NRS}). To define the generating function one needs to choose symplectic potentials for both forms,
$\Omega_{hom}$ and $\Omega_G$. 

There are two natural ways to define these potentials; the first way uses the set of coordinates based on
a given triangulation $\Sigma$.  The second way uses the  set of coordinates associated to the  odd part $H_-$ of 
$H_1(\Ch)$. The latter definition is more convenient in our approach to WKB expansion of the generating function
and relating it to the Bergman tau-function.  The former definition allows an easier description of the transformation properties of the generating function under the change of the underlying data.

In the next two sections we outline these two approaches.

\subsubsection{Definition of generating function using homological extension of shear coordinates and symplectic basis 
in $H_-$}

In view of  (\ref{defsym}) the natural choice of the potential $\theta_{hom}$ satisfying $\d \theta_{hom} =\Omega_{hom}$ 
 is
\be
\theta_{hom}=\sum_{j=1}^{g_-} \left(B_k \d A_k- A_k\d B_k\right)
\la{hompot}
\ee
where, as in  (\ref{defsym}), $A_k$ and $B_k$ are integrals of $v$ over cycles $(a_k^-,b_k^-)$ in $H_-(\Ch)$ (\ref{homco}).

 Let us discuss now the choice of potential for  Goldman's symplectic form.
 As it is shown in the App. \ref{monre}, the cover $\Ch_\Sigma$ is
 topologically and holomorphically equivalent to $\Ch$ irrespectively of the choice of the triangulation $\Sigma$ for given vertices $z_j$ (and the dual tri-valent graph $\Sigma^*$ with fixed vertices $x_j$).
 
  Choosing the same set of generators $(a_k^-,b_k^-)$ in $H_-(\Ch)$ as in (\ref{hompot}) we can write the form $\Omega_G$ in terms of the homological shear coordinates     $\FG_{a_j^-}$, $\FG_{b_j^-}$         (\ref{shearperiods}) as in  (\ref{GolDar}):
$$
\Omega_{G}= 2 \sum_{j=1}^{g_-} \d \FG_{a_j^-} \wedge \d \FG_{b_j^-}\;.
$$
%where $2g_-={\rm dim} H_-$.

The symplectic potential $\theta_G$ for the form  $-\Omega_{G}$ can be chosen as follows:
\be
\theta_{G}= \sum_{j=1}^{g_-} (\FG_{b_j^-}  \d \FG_{a_j^-}- \FG_{a_j^-}  \d \FG_{b_j^-})\;.
\la{thG}\ee

Now we are in a position to define the generating function $\Gcal$:
\begin{definition}
%\red{equivalent for all!!}
%triangulation $\Sigma$ such that the corresponding covering $\Ch_{\Sigma}$ 
 %is topologically equivalent to the  covering $\Ch$ given by $v^2=Q$. 
 
The generating function $\Gcal$ on $\Qcal_{g,n}[{\bf r}]$ (the "Yang-Yang" function according to the terminology of \cite{NRS}) is locally defined by
\be
\d \Gcal=\Fcal^* \theta_G- \theta_{hom}
\la{Gdef}
\ee
where $\theta_{hom}$ is given by (\ref{hompot}) and $\theta_G$ is given by (\ref{thG}).
 The homological shear coordinates are defined using the same  subset   $(a_j^-,b_j^-)$  of  generators  of  $H_-$ as the one used to define periods $(A_j,B_j)$.

\end{definition}

As it was shown in \cite{BK2iso}, in the framework of Fuchsian systems on the Riemann sphere the
generating function of the monodromy symplectomorphism can be identified with the isomonodromic tau-function.
%In present context the isomononodromic deformation is generically impossible by dimension count.

%One can associate many different generating functions to a given symplectic map; the difference is in the choice of symplectic potentials  on both sides. Our choice of symplectic potentials (\ref{thG}) for the Goldman symplectic form and (\ref{hompot}) for the homological symplectic form is made for technical convenience and symmetry.
 
 To summarize, in our framework the function $\Gcal$ depends on the following data.
 \begin{itemize}
 \item
 The Torelli marking (the choice of the canonical basis of cycles) of $\CC$ which determines the Bergman projective connection $S_B$.
 Let the basis  $(a_j,b_j)$, $j=1,\dots,g$ be related to a new basis  $(a^\s_j,b^\s_j)$  by a symplectic matrix $\s$:
 \be
\s=\left(\ba{cc} C & D \\ A & B \ea\right)\;:\hskip0.8cm
\left(\ba{c} b \\ a \ea\right)^\s=\s \left(\ba{c} b \\ a \ea\right)\;.
\la{symtr}
\ee
The Bergman projective connection then transforms as follows:
\be
S_B^{\sigma}= S_B-12 \pi i \sum_{j,k=1}^g \f{\p {\rm det}(C\Omega+D)}{\p\Omega_{jk}} v_j v_k
\ee
where $\Omega$ is the period matrix of $\Ccal$.

In the holomorphic case, when $n=0$,  the homological symplectic form (\ref{defsym}) 
coincides with the canonical symplectic form on the cotangent bundle to $\Mcal_g$.
On the other hand, in the case when all $r_j=0$ i.e. the differential $Q$ is allowed to have only simple poles at $z_j$, the homological symplectic form coincides with the canonical symplectic form on $T^*\Mcal_{g,n}$.

In both of these simpler  cases the new and old generating functions (assuming that all the other data remain
the same) are related by (see Proposition 5.1 of \cite{BKN} and \cite{Kor})
\be
\Gcal^{\s}= \Gcal +6\pi i \,\log \,\det (C\Omega+D)
\la{GTOR}
\ee
%where $\Omega$ is the period matrix of $\Ccal$ corresponding to the Torelli marking $\a$.
We expect this relation to extend in the same form to the case when $r_j\neq 0$.
%\red{??? At the moment we do not know how to relate the homological symplectic structure on $\Qcal_{g,n}[{\bf r}]$
%to the canonical structure on $\Mcal_{g,n}$. Therefore, we do not have a complete proof of validity of relation 
%(\ref{GTOR}) for $r_j\neq 0$, although we do know that it holds in the limit $r_j\to 0$, and conjecture that it remains valid for $r_j\neq 0$.}

 \item
 The choice of generators $(a_j^-,b_j^-)$ in $H_-$ which are used to define homological coordinates $(A_j,B_j)$ via (\ref{homco}) and homological shear coordinates $(\FG_{a_j^-},  \FG_{a_j^-})$ via (\ref{shearperiods}).
 Obviously, under the transformation 
 \be
\left(\ba{c} b_- \\ a_- \ea\right)^\s=\s_- \left(\ba{c} b_- \\ a_- \ea\right)  \;;\hskip0.8cm  \s_-=\left(\ba{cc} C_- & D_- \\ A_- & B_- \ea\right)\in Sp(2g_-,\Z)\;:\hskip0.8cm
\la{sym}
 \ee
 both symplectic potentials $\theta_G$ (\ref{thG}) and $\theta_{hom}$ (\ref{hompot}) remain the same.
 Therefore, $\Gcal$ also remains  invariant.
 
 \item
Finally, the  choice of triangulation $\Sigma$ used to define the shear coordinates $\zeta_e$. The definition of  homological linear combinations $(\FG_{a_j^-},  \FG_{a_j^-})$ also implicitly gets a non-trivial dependence on 
the choice of $\Sigma$ via $\zeta_e$. We expect that under an elementary Whitehead move $\Gcal$ transforms via a combination of dilogarithms.
%(which define the potential $\theta_G$) uses 
%also the choice of branch cuts (i.e. the perfect matching of vertices of $\Sigma^*$). 

%The dependence of $\Gcal$ on the choice of triangulation $\Sigma$ is more subtle. Namely, since any two triangulations of $\CC$ are related by a sequence of flips of diagonals; on the level of dual three-valent graphs $\Sigma^*$ such sequence corresponds to  a sequence of Whitehead moves. 
%Under an elementary Whitehead move the potential $\theta_{hom}$ does not change while the potential $\theta_G$ (\ref{thG}) does, and its exact transformation is not completely clear at the moment.

%We expect that the transformation of $\Gcal$ also involves a combination of corresponding dilogarithms;
%to get an exact relation one needs to find an  exact universal link between symplectic potentials (\ref{altpot}) below and 
%(\ref{thG}).

 \end{itemize}

To get the simplest transformation of the generating function under the choice of triangulation one can slightly modify the choice of symplectic potentials (and, therefore, the generating function itself) as shown in the next section.

\subsubsection{Definition of generating function using shear coordinates and periods around edges of $\Sigma^*$}

%Another version  of  symplectic potentials for both forms $\Omega_{hom}$ and $\Omega_G$ 
%can be proposed using the triangulation $\Sigma$ and the dual graph $\Sigma^*$.
%This leads to another version, $\tilde{\Gcal}$,  of the generating function which is closely related to $\Gcal$ and 
%is easier to study in some respects; however, $\Gcal$ is more suitable from the point of view of WKB formalism used in this paper.

Let us fix a  triangulation $\Sigma$ of $\CC$  and choose the branch cuts along all edges of 
$\Sigma^*$; then  the resulting covering is equivalent to the canonical cover $v^2=Q$ (see Prop.\ref{propSigma}). The form 
$\Omega_{hom}$ can be alternatively written in terms of periods of $v$ along cycles $\ell_e$ 
(the cycle $\ell_e$  goes around the edge $e^*$ of the graph $\Sigma^*$) as follows  \cite{BK1}:
\be
\Omega_{hom}=    \f{1}{4} \sum_{v\in V(\Sigma)} \sum_{e,e'\perp v,\;e'\prec e} \d \Pcal_{\ell_e'} \wedge \d \Pcal_{\ell_e}
\la{Golhom}
\ee
where  the notation $e\perp v$ means that the edge $e$ has $v$ as one of its endpoints;  $e'\prec e$ for two edges ending at $v$ means that 
the edge $e'$ precedes the edge $e$ under counter-clockwise ordering starting from a chosen "cilium" attached to the vertex $v$; 
$$
\Pcal_{\ell_e} =\int_{\ell_e} v\;.
$$
We choose  the  symplectic potential  (such that $\d\tilde{\theta}_{hom}=\Omega_{hom}$) as follows:
\be
\tilde{\theta}_{hom}=\f{1}{8}\sum_{v\in V(\Sigma)} \sum_{e,e'\perp v,\;e'\prec e}
( \Pcal_{\ell_e'}  \d \Pcal_{\ell_e}  - \Pcal_{\ell_e}  \d \Pcal_{\ell_e'})\;.
\la{altpot}
\ee

Let us now  consider  the (logarithmic) shear coordinates $\zeta_e$
($e\in E(\Sigma)$) on the character variety.
These coordinates go back to Thurston and Penner; see \cite{GMN,BK2} for  parametrization of monodromy matrices in terms of them.  The 
Goldman form on the symplectic leaf $CV_{g,n}[{\bf r}]$  looks as follows  (see (7.14) of \cite{BK1} and the recent paper \cite{Chekhov}. The convention is such that $\Omega P= -I$ i.e. the Poisson structure $\{p,q\}=1$ gives the symplectic form $\d p\wedge \d q$):
\be
\Omega_{G}=\sum_{v\in V(\Sigma)} \sum_{e,e'\perp v,\;e' \prec e} \d \zeta_{e'} \wedge \d \zeta_{e}\;.
\la{Gol1}
\ee
The  symplectic potential $\tilde{\theta}_G$  (such that $\d \tilde{\theta}_G=-\Omega_G$) associated to the representation (\ref{Gol1}) 
 can be chosen as follows:
\be
\tilde{\theta}_G= -\f{1}{2}\sum_{v\in V(\Sigma)} \sum_{e,e'\perp v,\;e'\prec e}
( \zeta_{e'}  \d \zeta_{e}- \zeta_{e}  \d \zeta_{e'})\;.
\la{altpotG}
\ee

Then  the alternative
 generating function $\tilde{\Gcal}$ is defined by the equation
\be
\d \tilde{\Gcal}=\Fcal^* \tilde{\theta}_G-\tilde{\theta}_{hom}\;.
\la{deftri}
\ee

The comment about transformation of the generating function $\Gcal$ (\ref{Gdef}) under the change of Torelli marking of $\CC$ is
also applicable to $\tilde{\Gcal}$: we expect that it transforms in the same way as (\ref{GTOR}) although no rigorous proof exists at the moment.

The transformation of $\tilde{\Gcal}$ under a change of triangulation $\Sigma$ is more explicit than the
transformation of $\Gcal$. 
Namely, under the Whitehead move on the edge $e$ the potential $\tilde{\theta}_G$ transforms as follows
(see Prop.7.1 of \cite{BK2}):
$
\tilde{\theta}_G\to \tilde{\theta}_G- 2 \d  L\left(\f{\kappa_e}{1+\kappa_e}\right)
$
where $L$ is the Rogers' dilogarithm and $\kappa_e=e^{2\zeta_e}$.
Similarly, one can derive the transformation law of the potential $\tilde{\theta}_{hom} $ (\ref{altpot})
 using  the requirement that periods $\Pcal_{\ell_e}$ corresponding to the edge $e$ 
and four adjacent edges $e_1,\dots,e_4$  transform  to preserve the pair $(\CC,Q)$.
This leads to the explicit transformation formula for the generating function $\tilde{\Gcal}$ under the Whitehead move.
We don't give here more details since in this paper we do not focus on  the global properties of the generating functions.

\section{Bergman tau-function over spaces of quadratic differentials with second order poles}
\la{tausec}

Here we review the notion of the Bergman tau-function on moduli spaces of quadratic differentials with
second poles and fixed biresidues.

The Bergman tau-function $\tau_B$ on moduli spaces appear in various context - from isomonodromy deformations to spectral geometry, Frobenius manifolds and random matrices, see the review \cite{KorRev}. 
In the context of moduli spaces of quadratic differentials with second order poles the Bergman tau-function was discussed in detail in \cite{BK1}, Section 4.1, where it is denoted by $\tau_+$; in  \cite{BK1} the
tau-function was considered on a stratum in the space of meromorphic quadratic differentials with fixed
biresidues.

Here we present previously known results about the Bergman tau-function on
moduli spaces of meromorphic quadratic differentials; the   equations for $\tau_B$ with respect to biresidues of $Q$ given here are new.
%Here we need to extend the 
%framework of \cite{BK1} by considering the dependence of $\tau_B$ on residues.
\subsection{Distinguished local coordinates on $\CC$ defined by $Q$}

 The divisor of $Q$ on $\CC$ looks as follows:
\be
(\qd)=\sum_{j=1}^N x_j-2\sum_{k=1}^n z_k= \sum_{i=1}^{n+N} d_i q_i\;.
\la{divQ}\ee

To define the set of {\it distinguished} coordinates on $\CC$ defined by the quadratic differential $Q$ 
we consider first the canonical cover $\Ch$ defined by $v^2=Q$. The divisor of abelian differential $v$ on $\Ch$ is given by
\be
(v)=\sum_{j=1}^N 2 x_j-\sum_{k=1}^n (z^{(1)}_k+ z^{(2)}_k)\;.
\la{divv}
\ee
The zeros $x_j$ of $v$ on $\Ch$ we denote by the same letters as zeros of $Q$ on $\CC$ since they are in one-to-one correspondence. On the other hand, $\pi^{-1}(z_j)=\{z_j^{(1)},z_j^{(2)}\}$. If 
$$
{\rm bires}|_{z_j} Q= r_j^2
$$
we can define $r_j$ such that 
\be
\Re( r_j) <0 
\ee
and
$$
{\rm res}|_{z_j^{(1)}} v = r_j\;,\hskip0.7cm 
{\rm res}|_{z_j^{(2)}} v = -r_j\;.
$$

Near  the zero $x_j$    the distinguished local coordinate on $\CC$ is defined by
\be
\zeta_j(x)=\left[\int_{x_j}^x v\right]^{2/3}\;,\hskip0.4cm j=1,\dots, N\;.
\la{dist1}
\ee
Near the double pole $z_j$  the distinguished local coordinate is defined  by
\be
\zeta_{N+j}(x)=\exp\left\{\f{1}{r_j}\int_{x_1}^x v\right\}\;,\hskip0.4cm j=1,\dots, n
\la{xijint}
\ee
where $x_1$ is a chosen zero of $\qd$.

To define the local coordinates $\zeta_{N+j}$ uniquely we connect poles $z_1,\dots,z_n$ by a system of arcs forming a tree graph $G$ on $\CC$. Furthermore, the first zero $x_1$ is connected to the first pole $z_1$ by 
a contour $\ell$. We denote the tree graph obtained by the lift of the union of $\ell$ and $G$ to $\Ch$ by 
$\Gh=\pi^{-1}(\ell\cup G)$.

The contours representing canonical cycles on $\Ch$ are assumed to not intersect the graph $\Gh$. Consider the fundamental polygon $\Ch_0$ of $\Ch$. Then the local coordinates $\zeta_{N+j}$ are uniquely defined in $\Ch_0\setminus \Gh$.

% \red{and the path of integration lies within the disk $D^{(1)}\subset \Ch$???}

\begin{figure}[htb]
\begin{center}
\includegraphics[width=8cm,angle=0]{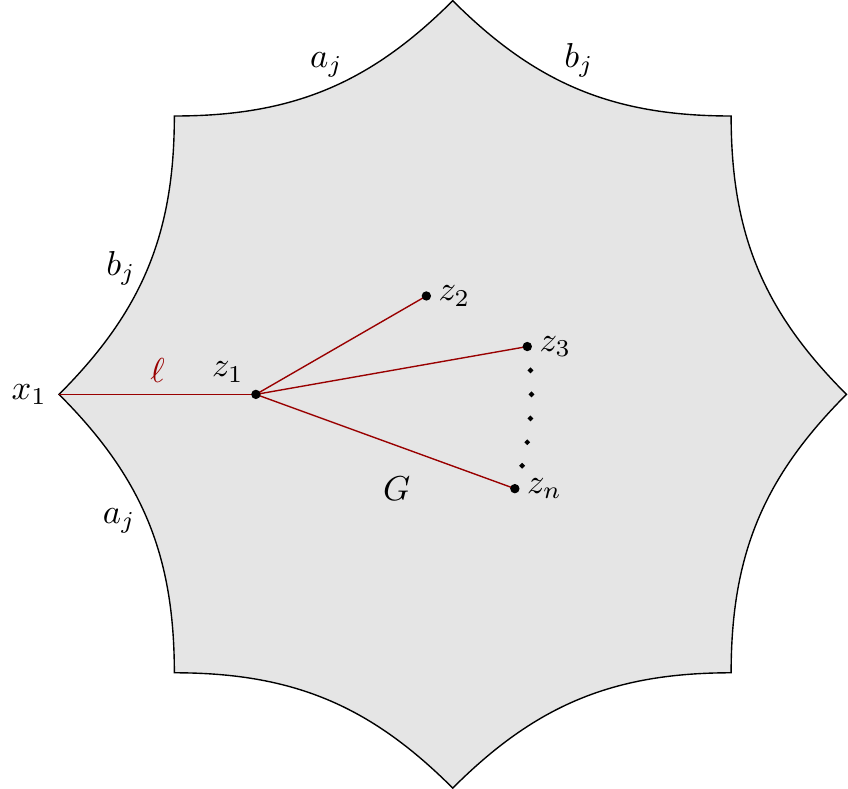}
\end{center}
\caption{Tree $G$ and contour $\ell$ within the fundamental polygon of $\CC$. }
\label{tau_cont}
\end{figure}

%To make the paths within the disk uniquely defined we choose a system of cuts $[x_1, z_j]$ which lie within the disk $D$ and assume that the paths of integration do not cross these cuts. 

\subsection{Definition of  $\tau_B$ and its properties}

%We start from explicit formula for $\tau_B$.
Denote by $E(x,y)$ the prime-form on $\CC$, by $\Acal_x$ the Abel map with  base-point $x$  and by $K^x$ the vector of Riemann constants (see e.g. \cite{Fay73}). 
For $n\geq 1$ the fundamental polygon of $\CC$ can always be chosen so that (the proof of this fact is parallel to  
the proof of Lemma 6 of \cite{JDG})
\be
\f{1}{2}\Acal_x((\qd))+2K^x=0\;.
\la{AK}\ee

Introduce the following multi-valued $g(1-g)/2$ - differential $c(x)$ \cite{Fay92}
\be
c(x)=\f{1}{W(x)}\left(\sum_{i=1}^g v_i(x)\f{\p}{\p w_i}\right)^g\theta(w,\Omega)\Big|_{w=K^x}
\ ,\qquad \ W(x):= \det \left[\frac {{\rm d}^{k-1}}{{\rm d} x^{k-1}} v_j\right]_{1\leq j,k\leq g}
\la{defCint}
\ee
where $\Omega$ is the period matrix of $\CC$, $\{v_j\}_{j=1}^{g}$ are holomorphic 1-forms on $\CC$ normalized by
$\oint_{a_i} v_j=\delta_{ij}$  and $\theta$ is the corresponding theta-function.

\begin{definition}\cite{JDG,contemp,BK1,KalKor}
Let the fundamental polygon of $\CC$  be chosen such that (\ref{AK}) holds. For a given Torelli marking and  the tree graph $\Gh$
the Bergman tau function $\tau_B$  is then defined  in terms of the divisor $(Q)$ (\ref{divQ}) 
 by the formula
\be
\tau_B=
c^{2/3}(x)\left(\f{\qd(x)}{\prod_{i=1}^{n+M}E^{d_i}(x,q_i)}\right)^{(g-1)/6} 
\prod_{i<j} E(q_i,q_j)^{\f{d_i d_j}{24}}\, .
\la{taupint}
\ee

The  prime-forms in (\ref{taupint})  are evaluated at the points $q_i$ as follows:
\be
E(x,q_i)=\lim_{y\to q_i} E(x,y) \sqrt{d\zeta_i(y)},
\la{defEpi}
\ee
\be
 E(q_i,q_j)=\lim_{x\to q_j, y\to q_i} E(x,y) \sqrt{d\zeta_i(y)} \sqrt{d\zeta_j(x)}
\la{defEppi}
\ee
where $\zeta$ are distinguished local coordinates (\ref{dist1}), (\ref{xijint}).
\end{definition}

The Bergman tau-function has the following properties:
\begin{itemize}
\item
The expression (\ref{taupint}) is constant with respect to $x$ although it seems to depend on  it \cite{JDG}. 

\item
The expression (\ref{taupint}) depends on the choice of the ``first'' zero  $x_1$ and on the  
integration paths between $x_1$ and a neighbourhood of $z_j$ in (\ref{xijint}).
The paths $[x_1,z_j]$ are chosen in the complement of the tree graph $\Gh$  (see Fig.\ref{tau_cont});
the change of the graph $\Gh$ within the fundamental polygon $\Ch_0$ affect the coordinates 
 $\{\zeta_{N+j}\}$ by a factor of the form
$$
\exp\left\{2\pi i\sum_{j,k} n_{jk}\f{r_j}{r_k}\right\}
$$
for some integers $n_{jk}$ \cite{KalKor}. We notice that this factor depends only on the residues $r_j$.

\item
Under the change of Torelli marking on $\CC$ with a matrix $\left(\ba{cc} A & B \\ C & D\ea\right)\in Sp(2g,\Z) $ the function  $\tau_B$  transforms as follows \cite{contemp,BK1,KorRev}:
\be
\tau_B\to \epsilon \, {\rm det}(C\Omega+D)\; \tau_B
\ee 
where $\epsilon^{48}=1$.
\item  
The  expression
\be
\tau_B^{48}\prod_{k=1}^n (d\zeta_{N+k}(z_k))^4
\la{defTpmint}
\ee
is invariant under the choice of local parameters  near $z_k$ \cite{BK1,KorRev}.
% and also under  the choice of signs in the definition of $v$.

\item
The function $\tau_B$ satisfies the following homogeneity property  (see \cite{KorRev}, Sec.6):
\be
\tau_B (\CC,\kappa Q)= \kappa^{   \f{5(2g-2+n)}{72}   } \tau_B (\CC, Q)\;.
\la{hom}
\ee

Defining the Euler vector field via
\be
E=\sum_{j=1}^{{3g-3+n}}  \left( A_j\f{\p}{\p A_j}+ B_j\f{\p}{\p B_j}\right)+ \sum_{j=1}^{{n}} r_k \f{\p}{\p r_k}
\la{EVF}
\ee
we get from (\ref{hom}):
\be
E\log \tau_B= \f{5(2g-2+n)}{72}\;.
\la{Etau}\ee
\end{itemize}

\subsection{Differential equations for $\tau_B$}

%We remind that the  set of local homological coordinates on the  moduli space of quadratic differentials with $n$ second order poles at $z_1,\dots, z_n$ is constructed as follows. One considers the homology group 
%$H_1(\Ch\setminus \{z_j^{(1)},z_j^{(2)}\}_{j=1}^n)$ and decomposes it as $H_+\oplus H_-$. 
%The generators in $H_-$ can be chosen  as in (\ref{homba}):
%\be
%\{s_j\}_{j=1}^{2g_-+n}= \{a_i^-,b_i^-\}_{i=1}^{3g-3+n},\{t_l^-\}_{l=1}^n
%\la{homba1}\ee
%where  $ g_-=3g-3+n$ and   the cycles $a_i^-,b_i^-,t_l^-$ are given by (\ref{abm1}), (\ref{abm}), (\ref{tjm}).

The group dual to  $H_1(\Ch\setminus \{z_j^{(1)},z_j^{(2)}\}_{j=1}^n)$  is the relative homology group 
$H_1(\Ch, \{ z_j^{(1)},z_j^{(2)}  \}_{j=1}^n)$
which we decompose into $H_+^*\oplus H_-^*$.
The space $H_-^*$ is dual to $H_-$ with respect to the intersection pairing. 
The set of generators of $H_-^*$ dual to generators (\ref{homba}) of $H_-$ is given by
\be
\{s_j^*\}_{j=1}^{2g_-+n}= \{-b_i^-,a_i^-\}_{i=1}^{g_-}\;,\;\;\{\kappa_l^-\}_{l=1}^n
\la{sstar}
\ee
where $\kappa_l^-$ is ($1/2$ of) the contour going from  $z_l^{(1)}$ to $z_l^{(2)}$ such that it does not intersect 
other contours from the basis (\ref{sstar}).  

We assume that the graph $\Gh$ lies entirely within the fundamental polygon of $\Ch$ 
assoiciated to the canonical basis (\ref{mainbasis}).
Moreover, we assume that  choice of contours $\kappa_l^-$  agrees with the system of cuts         $\Gh=\pi^{-1}(\ell\cup G)$            (where $G$ is a tree graph with vertices at $z_j$, see Fig.\ref{tau_cont}):   $\kappa_l^-$ equals to $1/2$ of the contour which starts at $z_l^{(1)}$, then goes to $x_1$ along cut shown in  Fig.\ref{tau_cont}, and then goes to $z_l^{(2)}$ on the second sheet along the contour which has the same projection to $\CC$.

Consider the set of homological coordinates on $\Qcal_{g,n}$ defined by (\ref{homco}).
%

%We have
%\be
%s_j^*\circ s_k=\f{\delta_{jk}}{{2}}\;.
%\la{sss}
%\ee

The equations   for $\tau_B$ with respect to the periods of $v$ along cycles (\ref{homba}) are given by the following theorem:

\begin{theorem} For a given Torelli marking denote by $S_B$ the Bergman projective connection on $\CC$ and define   the Bergman tau-function $\tau_B$ by (\ref{taupint}).
Then $\tau_B$ satisfies the following equations
\be
\frac{\p\log\tau_B}{\p A_i }=\frac{1}{24\pi i }\int_{b_i^-}\frac{S_B-S_v}{v} \;,\hskip0.7cm
\frac{\p\log\tau_B}{\p B_i}=-\frac{1}{24\pi i }\int_{a_i^-}\frac{S_B-S_v}{v}
\la{defBerg}
\ee
for $i=1,\dots,3g-3+n$ and
\be
\frac{\p\log\tau_B}{\p (2\pi i r_j) }=\frac{1}{24\pi i }\int_{\kappa_j^-}\left(\frac{S_B-S_v}{v}+\f{1}{2r_j^2}v\right)
\la{eqr}
\ee
for $j=1,\dots,n$ where the meromorphic projective connection $S_v$ is defined by (\ref{Sv}).
\end{theorem}

The proof of equations (\ref{defBerg}) is parallel to  the proof of equations for the Bergman tau-function on the space 
of holomorphic quadratic differentials with simple zeros, see \cite{contemp}. This proof of \cite{contemp} is obtained by reduction 
of the corresponding equations on the moduli spaces of holomorphic abelian differentials derived in \cite{JDG}.
Similarly, the formulas (\ref{defBerg}) are derived from equations for the Bergman tau-function on moduli spaces of 
meromorphic differentials of third kind obtained in \cite{KalKor}.
The equations with respect to residues (\ref{eqr}) are not used in this paper; they can be obtained 
from the  equations on the moduli spaces of holomorphic quadratic differentials by degeneration of the 
base curve $\CC$.

Therefore,  the differential of $\log\tau_B$ on the symplectic leaf $\Qcal_{g,n}[{\bf r}]$ is given by
the following expression:
\be
\d\log\tau_B\big|_{{\bf r}}=-\frac{1}{24\pi i }\sum_{j=1}^{{3g-3+n}} \left[\left(\int_{a_i^-}\frac{S_B-S_v}{v} \right)   \d B_j -  \left(\int_{b_i^-}\frac{S_B-S_v}{v} \right)   \d A_j  \right]\;.
\la{dlogtau1}
\ee

%On the other hand, on the full space  $\Qcal_{g,n}$ we have
% \be
%\d\log\tau_B=\d\log\tau_B\big|_{{\bf r}}+\frac{1}{24\pi i }\sum_{j=1}^{n}
%H_j\d r_j
%\la{dlogtau3}
%\ee
%where 
%$$
%H_j=\int_{\kappa_j^-}\left(\frac{S_B-S_v}{v}+\f{v}{2 r^2}\right) \;.
%$$

\section{WKB expansion  of the generating function of monodromy symplectomorphism}
\la{WKBsec}

%\subsection{Fock-Goncharov coordinates and periods of $s_j$}

%\vskip3.0cm

%\section{  WKB expansion of  Yang-Yang function }

%\subsection{Yang-Yang generating function of monodromy symplectomorphism: definition}

\subsection{Local connection problem: the Airy Stokes' phenomenon}
\la{Airysec}

In the neighbourhood of a zero of $Q$ (called  "turning point" in the theory of WKB approximation), there exists a change of coordinate $\zeta(z;\hbar)$ (see \cite{KT05}, Ch. 3) which is a formal analytic series in $\hbar$ and transforms the equation into the standard Airy equation 
\be
\label{Airy}
\hbar^2 \varphi''(x) - x \varphi(x)=0\;.
\ee
One of the features of this equation is its $\Z_3$ symmetry: if $\varphi(x)$ is a solution then
$\varphi(xe^{2\pi i/3})$ is also a solution. The point $x=0$, which is the  zero of the potential, is  the turning point 
of (\ref{Airy}).

The Stokes phenomenon  of the Airy equation near the point $x=0$ 
can be used to model the Stokes phenomenon for the general equation (\ref{Sint}) on a Riemann surface near a zero $x_j$ of the quadratic differential $Q$ (notice that  the 
Bergman projective connection $S_B$ vanishes in the coordinate $x$ on the complex plane and, therefore, the zero of $Q$ coincides with the zero of the potential). 

We summarize the Stokes phenomenon, also known as  the connection problem \cite{Wasow} or
{\it Airy parametrix} (see \cite{DKMVZ}).

%\red{One can then see (at least heuristically\footnote{To show that this is not a heuristic approach, one should show that the change of coordinate is actually analytic in a suitable sector in the $\hbar$ plane. This is not within our scope.}) that the local connection problem near the turning point becomes the familiar connection problem for the Airy equation, which we now report (see \cite{Wasov}). In a different branch of the literature on Riemann--Hilbert problems, this goes under the name of "Airy parametrix" (see \cite{DKMVZ}).}

There exist two special formal asymptotic expansions for a solution of (\ref{Airy})  \cite{AbrSteg}
\be
\label{AiryFormal}
f_{\pm}(x) \sim \frac{e^{\pm \f{i\pi}{4}} e^{\pm \frac{2}{3\hbar}x^{\frac{3}{2}}}}{
%2\sqrt\pi\,
x^{1/4}} \left[ \sum_{n=0}^{\infty} \frac{(\pm 1)^n \hbar^n\Gamma(n+\frac{5}{6})\Gamma(n+\frac{1}{6})\left(\frac{3}{4}\right)^n}{2\pi n! \, x^{3n/2}} \right]\;.
\ee
%These  formal series are equal   to the WKB expansions by taking the formal power series in $\hbar$ of the exponential of $s_{odd}$. 
The prefactor $e^{\pm \f{i\pi}{4}}$ is added here to simplify the form of the jump matrices below.
In the notation of  the previous section we have $Q=x(dx)^2$ and $v=\sqrt{x} dx$ i.e. the canonical covering is the two-sheeted cover of the Riemann sphere with branch points at $0$ and $\infty$.

One can write the  asymptotic expansion (\ref{AiryFormal}) in the form
%(an exercise using the recurrence relations for the expansion of $S$)
\be
f_\pm (x) =\exp \le[\pm \f{i\pi}{4}+
\int_0^x s(x;\pm \hbar)\sqrt{x}\d x
\ri]
\la{fpm}
\ee
where  
 $ s(x;\hbar) = \sum_{k\geq -1} s_k(x) \hbar^k$ and $s_k(x)$ are singular at $x=0$. The antiderivative in understood in the following sense: for any power series centered at $0$ the integration is made term by term assuming that \blue{$\int_0^x x^k\d x=x^{k+1}/(k+1)$} for any $k\neq -1$ (this convention can be interpreted also as $1/2$ of the definite integral between two points, $x^-$ and $x^+$, lying on different sheets of the Riemann surface of the function $\sqrt{x}$) . Let  $s_{odd}$ contain only the odd powers of $s$ in $\hbar$. Then 
 (similarly to the link (\ref{evenodd}) between $s_{odd}$ and $s_{even}$ in the case of general equation on a Riemann surface derived below) we get
\be
f_\pm=\frac {e^{\pm \f{i\pi}{4}}}{\hbar^{1/2}(s_{odd}\sqrt{x})^{1/2}} \exp\le[ \pm \int_0^x s_{odd}\sqrt{x} \d x\ri]
\la{Phipm}
\ee
%with $ s(x;\hbar) = \sum_{\ell\geq -1} s_\ell(x) \hbar^ \ell$
where
\be
s_{-1} =  1\;,\hskip0.7cm  s_0 = - \frac 1 {4 x^{3/2}}\;, \hskip0.7cm 
s_k = -\frac {d_k} {2^{3k+2}} \frac1 {x^{\frac 3 2 k+\frac 5 2}} \;, \hskip0.7cm  \ell\geq 1\; .
\ee
The recurrence relations between $s_k$ obtained by substitution of (\ref{fpm}) into (\ref{Airy})  imply that the coefficients $d_k$ in this formula are  positive integers satisfying the recurrence relation
\be
d_0=1\;,\qquad d_{k+1} = (6k+4) d_k + \sum_{j=0}^k d_j d_{k-j}\;.
\ee 
The numbers $d_k$ form the sequence sequence number $A062980$ in OEIS.org, and we refer ibidem for interesting combinatorial interpretations (in particular, in the problem of counting of three-valent graphs on surfaces).

\paragraph{Stokes' phenomenon.}
The Stokes' phenomenon for the Airy equation has been known for more than a century. The standard Stokes' phenomenon refers to the discontinuous asymptotic behaviour of a solution of (\ref{Airy}) as $x \to\infty$ in different directions; this translates to a Stokes' phenomenon as $\hbar \to 0$ by rescaling $x\to \hbar^{-2/3} x$.

\begin{figure}[ht]
\begin{center}
\begin{tikzpicture}[scale=1.7]
\node [below right] at(0,0) {$0$};
\draw [red!90!black,postaction={decorate,decoration={markings,mark=at position 0.7 with {\arrow[line width=1.5pt]{<}}}}] (0,0) to
node [pos=0.6,right] {$U$} (120:1);
\draw [<->]   (0.1,0) arc  (0:120:0.1)node[above, pos=0.5]{$\frac {2\pi}3$};
\draw [red!90!black,postaction={decorate,decoration={markings,mark=at position 0.7 with {\arrow[line width=1.5pt]{<}}}}] (0,0) to
node [pos=0.6,right] {$U$} (-120:1);
\draw [blue,postaction={decorate,decoration={markings,mark=at position 0.7 with {\arrow[line width=1.5pt]{<}}}}] (0,0) to
node [pos=0.6,above] {$L$} (0:1);
\draw[decoration={aspect=0.5, segment length=02, amplitude=0.7,coil},decorate] (0,0) to (180:1);

\draw [line width =0, ,postaction={decorate,decoration={markings,mark=at position 0.7 with {\arrow[line width=1.5pt]{<}}}}] (180:0.5) to
node [pos=0.6,above] {$J$} (180:0.51);
\end{tikzpicture}
\end{center}
\caption{Standard configuration of contours and jump matrices in the Riemann-Hilbert formulation of the Stokes phenomenon for the Airy equation.}
\label{StAiry1}
\end{figure}
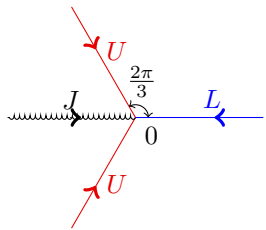

Classically, the Stokes' phenomenon for the Airy equation is described using the configuration of contours shown in
 Fig. \ref{StAiry1}. In each of the four sectors of the figure there is a {\it distinguished} basis 
 $(\varphi_{j,+},\;\varphi_{j,-})$ of solutions of \eqref{Airy} which is asymptotic (in the usual Poincar\'e sense) to the formal basis $(f_+,\,f_-)$ \eqref{AiryFormal} as $\hbar \to 0^+$; for a modern presentation of this phenomenon we refer to \cite{DKMVZ}.
 
 \vskip0.4cm
 
 The jump matrices on the four contours  indicated in Fig. \ref{StAiry1}  are given by  
 \be
 L = \le(\ba{cc} 1 & 0\\ -1 & 1 \ea\ri)\;,\hskip0.7cm
U = \le(\ba{cc} 1 & 1 \\ 0 & 1 \ea \ri)\;,\hskip0.7cm
J= \le(\ba{cc} 0 & -1 \\ 1  & 0 \ea \ri)\;.
\la{LUJ}\ee
 These are the Stokes' matrices, which indicate how the distinguished bases in different sectors are related one to the other. The notation is that, say,  the matrix $J$  on an oriented ray indicates  the relationship $\Phi_{left} = \Phi_{right} J$  where $\Phi_{left}$ is the basis of solution on the sector to the  left of the  ray and $\Phi_{right}$ on the right of the ray. Note that $LUJU=\1$, which indicates  that the Airy equation has no monodromy (since its only  singularity is at $x=\infty$).

All fractional powers appearing in the expression \eqref{AiryFormal} are defined via their principal determination.
% in the fractional powers in the exponent and for $x^{1/4}$. 
 With this choice, the jump of $x^{3/2}$  is placed along $\R_-$ so that the real part, $\Re (x^{3/2})$ is positive on $\R_+$ and negative on the two red rays.  The form of the matrix $J$ (\ref{LUJ}) can be 
easily seen from the definition  of the asymptotic expansions (\ref{Phipm}). Namely, $s_{odd}$ remains the same when
$x$ crosses $\R_-$, but $\sqrt{x}$ changes sign. Thus, essentially, $f_+$ is interchanged with $f_-$;
there is also an extra power of unity which comes from transformation $x^{1/4}\to e^{\pi i/2} x^{1/4}$. 
Therefore, the ``boundary value" (understood term-by term in $\hbar$-expansion) of $f_+$   from above of $\R_-$ is equal to the boundary value of $f_-$ from below of $\R_-$. Similarly, the boundary value of $f_-$  from above of $\R_-$ is equal to {\it minus}  the boundary value of $f_+$ from below of $\R_-$, in agreement with Fig.\ref{StAiry1}.

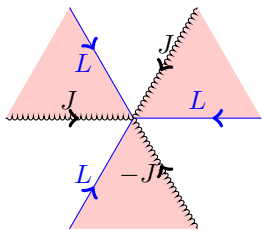
\begin{figure}[h]
\begin{center}
\begin{tikzpicture}[scale=1.7]

\fill[red!20!white] (0,0) to (60:1) to (0:1) to cycle;
\fill[red!20!white] (0,0) to (-120:1) to (-60:1) to cycle;
\fill[red!20!white] (0,0) to (180:1) to (120:1) to cycle;

\draw [blue,postaction={decorate,decoration={markings,mark=at position 0.7 with {\arrow[line width=1.5pt]{<}}}}] (0,0) to
node [pos=0.5,above] {$L$} (0:1);
\draw [blue,postaction={decorate,decoration={markings,mark=at position 0.7 with {\arrow[line width=1.5pt]{<}}}}] (0,0) to
node [pos=0.5,left] {$L$} (-120:1);
\draw [blue,postaction={decorate,decoration={markings,mark=at position 0.7 with {\arrow[line width=1.5pt]{<}}}}] (0,0) to
node [pos=0.5,left] {$L$} (120:1);
\draw [line width =0, postaction={decorate,decoration={markings,mark=at position 0.7 with {\arrow[line width=1.5pt]{<}}}}] (180:0.5) to
node [pos=0.6,above] {$J$} (180:0.51);
\draw [line width =0, postaction={decorate,decoration={markings,mark=at position 0.7 with {\arrow[line width=1.5pt]{<}}}}] (60:0.5) to
node [pos=0.6,above] {$J$} (60:0.51);
\draw[decoration={aspect=0.5, segment length=02, amplitude=0.7,coil},decorate] (0,0) to (180:1);\draw[decoration={aspect=0.5, segment length=02, amplitude=0.7,coil},decorate] (0,0) to (60:1);
\draw[decoration={aspect=0.5, segment length=02, amplitude=0.7,coil},decorate] (0,0) to (-60:1);
\draw [line width =0,postaction={decorate,decoration={markings,mark=at position 0.7 with {\arrow[line width=1.5pt]{<}}}}] (-60:0.5) to
node [pos=0.6,left] {$-J$} (-60:0.51);
\end{tikzpicture}
\end{center}
\caption{The Riemann-Hilbert problem on 6 contours obtained via transformation of the RH problem in Fig.\ref{StAiry1}. The Riemann surface of $\sqrt{x}$ here has three branch cuts meeting under equal angles $2\pi/3$ at $0$.}
\label{StAiry2}
\end{figure}

For the sake of symmetry we  re-formulate the RHP as shown in  Fig.\ref{StAiry2} by introducing two additional 
contours of discontinuity  $\arg x = \pm \pi/3$ and multiplying the basis $\Phi$  in the sector $\mathrm{arg}\, x\in (\pi/3,5\pi/3)$ by $J$. This creates the additional two jumps with jump matrices  $J$ and $J^{-1}=-J$, shown on  Fig.\ref{StAiry2}. In addition,   both jump matrices $U$ get conjugated by $J$ which produces the jump matrices $L$ on the rays ${\rm arg} \;x =\pm 2\pi/3$.

The new  jumps along   the rays ${\rm arg} \;x =\pm \pi/3$ are the same (up to a sign)  as the jump along  ${\rm arg} \;x =\pi$. This allows to interpret the rays ${\rm arg} \;x =\pm \pi/3$    as extra branch cuts for  $\sqrt{x}$. In each of the six sectors of Fig. \ref{StAiry2} the  distinguished basis of solutions is asymptotic 
to the formal basis $(f_+, f_-)$ provided that the $\sqrt{x}$ in (\ref{Phipm}) has not one, but three branch cuts coming to $x=0$ along the rays ${\arg x}=\pi, \pm \pi/3$. 

There is still a residual violation of the  $\Z_3$ symmetry in the Riemann-Hilbert problem in Fig. \ref{StAiry2} due to the extra sign in the jump matrix along ${\arg x}=- \pi/3$; this violation appears due to the term $x^{1/4}$ in (\ref{Phipm}).  In the sequel we restore the  $\Z_3$ symmetry  by treating jump matrices as elements of the  $\mathbb P SL(2)$ monodromy representation of (\ref{Sint1}).

\vskip0.2cm

The goal of  the next transformation is to obtain a Riemann-Hilbert problem with jumps only on the critical trajectories of the quadratic differential $Q=x(\d x )^2$. For that purpose we further rotate the the three branch cuts
of $x^{1/2}$ clockwise by $\pi/3$ so that now they run along the critical trajectories themselves. This amounts to multiplying by $\pm J$ the basis of solutions in the shaded regions of Fig.\ref{StAiry2}.

\begin{figure}[h]
\begin{center}
\begin{tikzpicture}[scale=1.7]
\draw [blue,postaction={decorate,decoration={markings,mark=at position 0.7 with {\arrow[line width=1.5pt]{<}}}}] (0,0) to
node [pos=0.6,above] {$A$} (0:1);
\draw [blue,postaction={decorate,decoration={markings,mark=at position 0.7 with {\arrow[line width=1.5pt]{<}}}}] (0,0) to
node [pos=0.6,right] {$-A\sim A$} (-120:1);
\draw [blue,postaction={decorate,decoration={markings,mark=at position 0.7 with {\arrow[line width=1.5pt]{<}}}}] (0,0) to
node [pos=0.6,right] {$-A\sim A$} (120:1);
\end{tikzpicture}
\end{center}
\caption{The Riemann-Hilbert problem on critical trajectories.
The symbol $\sim$ means  equivalence  of the jump matrices in $\mathbb P \mathrm{SL}_2$.
%\red{What happens on a Riemann surface? The branch cuts there go between zeros, not along critical trajectories...}
}
\label{StAiry3}
\end{figure}
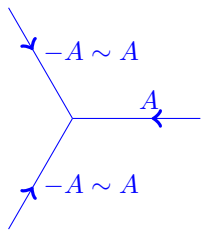
 
 The result of this operation is the Riemann--Hilbret problem depicted in Fig.\ref{StAiry3}, where 
\be
A = -LJ=\left( \ba{cc}  0 & 1 \\ -1 & -1 \ea   \right)\;.
\la{defA}
\ee
Although two of the jump matrices equal to $-A$, and one of them equals $A$, in the $PSL(2)$ group, 
all of these  matrices coincide. This makes the Riemann-Hilbert problem $\Z_3$ symmetric,
in agreement with the $\Z_3$ symmetry of the Airy equation itself. Notice that in this way we could 
reconstruct the original jump matrices from Fig.\ref{StAiry1} via ``reverse engineering".
 
% \br\rm The Riemann-Hilbert problems shown  in Fig.\ref{StAiry2}  and Fig.\ref{StAiry3} 
 %are explicitly $\Z_3$ symmetric if the jump matrices are understood as elements of  $PSL(2)$.
%In particular, the matrix $A$ (\ref{defA}) satisfies  $A^3=I$. This agrees with the $\Z_3$ symmetry of the Airy equation itself. The $\Z_3$ symmetry allows actually to deduce the form of the jump matrices $J$, $U$ and $L$ 
%(\ref{LUJ}) in an elementary way from the relation $A^3=I$ via ``reverse engineering".
%\er

%\br
%The choice of $\pm J$ in the transformation corresponds to  different possible choices of the branch of  $x^\frac 14$ and may lead to different orientations on the added green jumps in Fig. \ref{StAiry}. Since we are interested only in the monodromy representation in $\mathbb P SL_2$,  this issue is irrelevant. Moreover, for the same reason, the orientation on the green jumps is not important because $J^{-1} = -J$ and hence represents the same matrix in $\mathbb P SL_2$.
%\er

\subsection{Global WKB Riemann-Hilbert problem}
\la{WKBRH}

\subsubsection{Formal WKB expansion}

%The problem of understanding of the  generating function of the monodromy symplectomorphism (the Yang-Yang function) for equation (\ref{Sint}) was posed in \cite{NRS}.
%Here we briefly discuss the WKB expansion of the Yang-Yang function in the framework of recent results of \cite{Alleg} where the relationship between Voros symbols and the 
%complex shear coordinates (the $SL(2)$ case of the Fock-Goncharov coordinates) on the character variety was established.

%We shall follow the  conventions of \cite{BKN} and write the main equation as follows:
Let us now rescale the quadratic differential $Q$ by a large parameter and consider the equation of the form 
\be
\partial^2\phi+\left(\f{1}{2} S_B - \f{1}{\hbar^2} Q\right)\phi=0\;.
\la{eqhbar}
\ee
The parameter $\hbar$ then enters the equation of the corresponding canonical cover $\Ch_\hbar$:
\be
v_{\hbar}^2=\f{1}{\hbar^2} Q\;.
\la{covhbar}
\ee
The cover $\Ch_\hbar$ is of course conformally equivalent to the $\hbar$-independent  cover
$
v^2=Q
$
via the  rescaling $v={\hbar}v_\hbar$.

Choose some base point $x_0$ (later on $x_0$ is going to be chosen to be one of zeros of $Q$). Then 
in terms of the coordinate $z(x)=\int_{x_0}^x v$ and the function $\varphi(x)=\phi \sqrt{v(x)}$ we have
\be
\varphi_{zz}+(q(z)- \e^{-2})\varphi=0
\la{eqcoo}
\ee
where
$$
q=\f{\Scal_B-\Scal_v}{2v^2}\;.
$$

To study the limit $\e\to 0$ we introduce the asymptotic series   $s=\sum_{k=-1}^\infty \e^k s_k$  and
write the asymptotic series for solution of (\ref{eqhbar}) in the form
\be
f=v^{-1/2} \exp\left\{\int_{x_0}^x (\e^{-1} s_{-1}+s_0+\e s_1+\dots)v\right\}
\la{formWKB}
\ee
where $s_k$ are meromorphic functions on $\Ch$  and $x_0$ is a basepoint.
We introduce also the meromorphic differentials 
\be
v_k=s_k v\;.
\la{defvk}
\ee

The function $s$ satisfies the Riccati equation
\be
  \d s+ v s^2 =-qv + \e^{-2}v
  \la{Ric}
\ee
(notice that on both sides of this equation we have 1-forms)
or
$$
\d\left(\sum_{k=-1}^\infty \e^k s_k\right)+v\left(\sum_{k=-1}^\infty \e^k s_k\right)^2=-qv+\e^{-2}v\;.
$$
The coefficients of  $\e^{-2}$, $\e^{-1}$ and $\e^0$ give
$$
s_{-1}=\pm  1 \;, \hskip0.7cm s_0=0\;, \hskip0.7cm
 s_1=- q/2\; .
$$
so that $v_{-1}=\pm v$.

For all other coefficients we have 
\be
\d s_k+ v\sum_{j+l=k\atop j,l\geq -1}  s_j s_l =0, \ \ \ k>0\;.
\la{recur}
\ee
Thus
$$
 s_{k+1}=-\f{1 }{2s_{-1}}\left(\f{\d s_k}{ v}+\sum_{j+l=k,\atop j,l\geq 0} s_j s_l\right)\;.
$$

In particular, for $k=1,2$ we get 
\be
s_2= \f{ -\d s_1}{2 s_{-1} v}\;,
\ \ \qquad s_3=  \f{1}{s_{-1}}\le(\f{s_1^2}{2}-\f{1}{4v}\d\left(\f{\d s_1}{v}\right)\ri) \;.
\ee

%\begin{lemma}
Functions $s_{2k+1}$ are symmetric under $\mu$ 
%\blue{and also under $s_{-1}\mapsto -s_{-1}$}, 
while $s_{2k}$ are skew-symmetric. Therefore,  the 1-forms 
$
v_{2k}=s_{2k}v
$
 can be identified with  meromorphic 1-forms
on $\Ccal$, while $v_{2k+1}=s_{2k+1}v$ are meromorphic 1-forms on $\Ch$ anti-symmetric under $\mu$.

%\end{lemma}

%{\it Proof.} We proceed by induction. The induction step yields 
%\be
%s_{2k+1}=\f{-1}{2s_{-1}}\left(\f{\d s_{2k}}{v}+\sum_{j+l=2k\atop l,k\geq 0} s_j s_l\right)\;.
%\ee
%By induction hypothesis all products in the sum are even, as well as the first term; thus $s_{2k+1}$ is even under the involution $\mu$. 
%Similarly,
%\be
%s_{2k}= \f{-1}{2s_{-1}}\left(\f{\d s_{2k-1}}{v}+\sum_{j+l=2k-1\atop l,k\geq 0} s_j s_l\right)
%\ee
%and all the terms in the right-hand side are odd by induction hypothesis.
%\QED
 
Consider now the corresponding even and odd parts:
\be
s_{even}=\sum_{l=0}^\infty s_{2l} \e^{2l}\;, \hskip0.7cm
s_{odd}=\sum_{s=-1}^\infty s_{2s+1} \e^{2s+1}\;.
\ee
Below we shall assume $s_{-1}=1$ (i.e. $v_{-1}=v$); to get the second asymptotic series corresponding to the choice
$v_{-1}=-v$ it is sufficient to apply the involution $\mu$ and use the fact that $\mu^*v =-v$.
%and
%\be
%v_{even}=s_{even}v\;\hskip0.7cm 
%v_{odd}
%\Vdef=\hbar s_{odd}v
%\la{Vdef}
%\ee
Notice that $s_{even}(x^\mu)=-s_{even}(x)$ and  $s_{odd}(x^\mu)=s_{odd}(x)$ while
\be
\mu^*v_{even}=v_{even}\;,\hskip0.7cm
\mu^*v_{odd}=-v_{odd}\;.
\ee
\begin{lemma}
The following equation holds:
\be
\d s_{odd}= -2 v\, s_{even} s_{odd}\;.
\la{evenodd}
\ee
\end{lemma}
{\it Proof.} This directly follows from equations (\ref{recur}). Namely, the statement of the lemma is written as
\be
\sum_{k=-1}^\infty \e^{2k+1}\d s_{2k+1} = -2v \left(\sum_{l=0}^\infty s_{2l}\e^{2l}\right)\left(\sum_{j=-1}^\infty s_{2j+1}\e^{2j+1}\right)
\ee
which is equivalent to (\ref{recur}) (the factor of 2 appears since the sum in (\ref{recur}) goes twice over all pairs of indices). \QED

%$\red{??? According to Theorem  7.17 of \cite{Alleg}, the sum of V\"{o}ros symbols over a path between two zeros of $v$ lying in some rectangle bounded by critical trajectories of $v$ (understood in the sense of the Borel resummation scheme), tends  to the Fock-Goncharov coordinate corresponding to the opposite diagonal of the same rectangle in the limit $\hbar \to 0$. }

\subsubsection{Riemann-Hilbert problem}

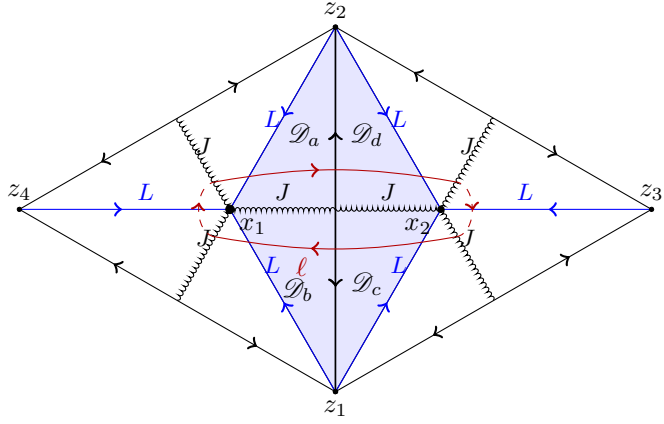
\begin{figure}[h]
\begin{center}
%\begin{tikzpicture}[scale=0.8]
%\TriangleCut{1}{1};
%\end{tikzpicture}
%\qquad
\begin{tikzpicture}[scale=1.4]
\begin{scope}[xshift=-28.5, rotate=60]
\draw [fill=white!90!blue] (0,0) to (-120:2) to (0:2) to cycle;
\node at (-15:1){$\scr D_a$};
\node at (-110:1){$\scr D_b$};
\TriangleCut{3}{1};
\draw[fill] (0,0) circle[radius =0.04];
\node[above] at(120:2) {$z_4$};
\node[below right] at(0,0) {$x_1$};
\end{scope}
\begin{scope}[xshift=28.5, rotate=-120]
\draw [fill=white!90!blue] (0,0) to (0:2) to (-120:2) to cycle;
\node at (-15:1){$\scr D_c$};
\node at (-105:1){$\scr D_d$};
\TriangleCut{3}{1};
\draw[fill] (0,0) circle[radius =0.03];
\node[below left] at(0,0) {$x_2$};
\node[below] at(0:2) {$z_1$};
\node[above] at(-120:2) {$z_2$};
\node[above] at(120:2) {$z_3$};
\end{scope}
 \draw [red!70!black , postaction={decorate,decoration={markings,mark=at position 0.6 with {\arrow[line width=1pt]{<}}}}] ($(0:1)+(60:0.3)$) 
 to [out=170,in=10]   ($(180:1)+(120:0.3)$);
 \draw [red!70!black , postaction={decorate,decoration={markings,mark=at position 0.6 with {\arrow[line width=1pt]{>}}}}] ($(0:1)+(-60:0.3)$) 
 to [out=190,in=-10]  node[pos=0.65, below] {$\ell$} ($(180:1)+(-120:0.3)$);

 \draw [dashed, red!70!black , postaction={decorate,decoration={markings,mark=at position 0.6 with {\arrow[line width=1pt]{<}}}}]($(0:1)+(-60:0.3)$)
 to [out=10,in=-10]    ($(0:1)+(60:0.3)$) ;
 \draw [dashed, red!70!black , postaction={decorate,decoration={markings,mark=at position 0.6 with {\arrow[line width=1pt]{>}}}}]($(180:1)+(-120:0.3)$)
 to [out=170,in=190]    ($(180:1)+(120:0.3)$) ;
\end{tikzpicture}
\end{center}
\caption{Two neighbouring faces of $\Sigma_Q$. Waved black curves are 
edges of $\Sigma_Q^*$; these are branch cuts which carry the jump matrix $J$. The edges of the critical graph $\Gamma_Q$ carry the jump matrix $L$. Edges of 
$\Sigma_Q$ (black) carry the diagonal jump matrices expressed via contour integrals of $s_{odd}v$. 
}
\label{faces}
\end{figure}

The graph $G_Q$ where the Riemann-Hilbert problem will be formulated is shown in Fig.\ref{faces}, \ref{FigTrian};
it is the union of the critical graph $\Gamma_Q$ of the differential $Q$ and the graphs $\Sigma_Q$
and $\Sigma_Q^*$:
\be
G_Q=\Gamma_Q\cup \Sigma_Q\cup \Sigma_Q^*
\la{defG}
\ee
The  edges of the graph are provided with the orientation and jump matrices indicated in Fig.\ref{faces}.
%The jump matrices on the edges of $\Gamma_Q$ equal to the matrix $L$ while the edges belonging to $\Sigma_Q^*$ come with jump matrix $J$. Finally, the jump matrices on the edges of $\Sigma_Q$ will be equal to diagonal matrices defined below and expressed via Voros symbols.

Let us explain how this kind of Riemann-Hilbert problem arises from the  WKB approximation.
 The formula (\ref{formWKB}) can be used to produce two 
asymptotic expansions for solutions of (\ref{eqhbar}) in each triangle face of the graph $\Sigma_Q$ shown in Fig.\ref{faces}. These asymptotic expansions  are obtained from    (\ref{eqhbar}) 
by expressing $s_{even}$ in terms of $s_{odd}$ via (\ref{evenodd}). In the $j$th face of $\Sigma_Q$ the ``initial point of integration" (in the sense which is clarified below)  is chosen to coincide with the zero $x_j$ of $Q$ contained in this face; thus the corresponding asymptotic expansions inherit the same index. The following expression is a
straightforward generalization of the formulas (\ref{Phipm}) 
\be
f^{(j)}_\pm=\frac {e^{\pm \f{i\pi}{4}}}{\hbar^{1/2} (v s_{odd})^{1/2}} \exp\le[ \pm \int_{x_j}^x  s_{odd}v\ri]
\la{Phipmj}
\ee
(notice that  $f^{(j)}_\pm$ are $-1/2$-differentials while in the case of Airy equation $f_\pm$ given by 
(\ref{Phipm}) are functions). 
The differential $v_{odd}=s_{odd} v$ is singular at $x_j$ but its residues at $x_j$ vanish due to skew-symmetry; therefore we can define $\int_{x_j}^x  s_{odd}v$ in
the same sense as the integral from $0$ in (\ref{Phipm}) \cite{KT05}. Namely, we  consider $x$ as $x^{(1)}$ i.e. the point on the first sheet of the canonical cover $\Ch$ (recall that, according to App.\ref{monre1} the first sheet consists of faces 
of $\Sigma_Q^*$ where the real parts of residues of $v$ at $z_j$ are negative). Then $x^{(2)}$ is the 
point on the second sheet and we define 
$$
 \int_{x_j}^x  s_{odd}v=\f{1}{2} \int_{x^{(1)}}^{x^{(2)}}  s_{odd}v
$$
where the integration contour follows the edge of $\Sigma_Q^*$ on both sheets.

%\red{ Need this? The $PS(2)$ monodromy group of equation (\ref{eqhbar}) can be defined as the monodromy group of the Schwarzian
%equation
%\be
%\Scal(F,\xi)= s_B-\f{2}{\hbar^2}q
%\la{Schwarzian}
%\ee
%where $S_B= s_B (\d\xi)^2$, $Q=q(\xi)(\d \xi)^2$ and $\xi$ is a local coordinate; $F=\phi_2/\phi_1$ is the  ratio of two solutions of (\ref{eqhbar}).
%The jump matrix $\left(\ba{cc} a & b \\c & d\ea\right)$ on a contour $l$ is then interpreted as the following relation between $F_{l}$ and $F_{r}$:
%\be
%F_{l}=\f{a F_{r} +b}{c F_{r} +d}
%\la{mons}
%\ee}

Now we are in a position to introduce the following definition.

\begin{definition}
\label{defJL}
\rm
The model $PSL(2)$ Riemann-Hilbert problem  has the following set of jump matrices
on edges of the graph $G_Q$ (\ref{defG}) as shown in Fig.\ref{faces} (see (\ref{LUJ}) for the definition of $L$ and $J$):
\begin{itemize}
\item
The matrix $J$ on black waved contours connecting zeros of $Q$ along edges of $\Sigma_Q^*$.
\item
The matrix $L$ on edges of the critical graph $\Gamma_Q$.
\item
Diagonal matrices depending on moduli on the edges of $\Sigma_Q$; on the edge $e$ connecting 
two poles of $Q$ the jump matrix is given by
\be
Z_e=\left(\ba{cc} e^{\zeta_e} & 0\\
0 & e^{-\zeta_e} \ea\right)\;;
\la{Ze}
\ee
the expression $e^{2\zeta_e}$ is called the "complex shear coordinate", which is the simplest case of the general Fock-Goncharov coordinates.
\end{itemize}
\la{RHpro}
\end{definition}

 The $PSL(2)$  Riemann-Hilbert problem described in Def.\ref{RHpro}, including the coordinates $\zeta_e$,  arises via the following mechanism.
 
Let us choose a local coordinate $\xi$ in the simply connected region consisting of four faces of the graph $G_Q$ bounded by four critical trajectories as shown in the shaded diamond region  in Fig. \ref{faces}.

 Note that each face $\scr D$ of $G_Q$ has exactly one critical trajectory of $\Gamma_Q$ on its boundary, one branch-cut and one zero $x_j$ of $Q$.

In a neighbourhood  of the critical trajectory issuing from $x_j$, the real part of $\int_{x_j}^x v$ is  positive  because of our choice of sheets. This means that in this neighbourhood, as $\hbar \to 0^+$, the distinguished solution $\varphi_-^{ {(\mathscr D)} }$ is {\it recessive} (i.e. tends exponentially to zero), while $\varphi_+^{{(\mathscr D)}}$ is {\it dominant} (i.e. explodes exponentially).
\begin{definition}
\label{def1}
The solution $\varphi_-^{(\scr D)}$ is the unique solution of \eqref{eqhbar} which is asymptotic to $f_-^{(j)}$ in the region $\scr D$ as $\hbar \to 0$.
\end{definition}
The reason why this solution is unique is that we cannot add any multiple of a dominant solution to it. 

To define uniquely $\varphi_+^{\scr D}$ let us assume that $\scr D$ is to the right of the branch cut (oriented towards the zero $x_j$, as in Fig. \ref{StAiry2}) and 
$ \wt {\scr D}$ be the unique face of $G_Q$ to the left of the same branch-cut (referring to   Fig. \ref{faces}
if  $\scr D$ is chosen to be $\scr D_a$ then $ \wt {\scr D}$ is $\scr D_b$).
Then we define 
\begin{definition}
\label{def2}
The solution $\varphi_+^{\scr D}$ is the unique analytic continuation to $\scr D$ of the solution $-\varphi_-^{\wt {\scr D}}$. Similarly the solution $\varphi_+^{\wt {\scr D}}$ is the unique analytic continuation to $\wt{\scr D}$ of the solution $\varphi_-^{ {\scr D}}$.
\end{definition}

The two Definitions \ref{def1}, \ref{def2} determine uniquely the {\bf distinguished} basis in each region $\scr D$ and imply that the jump matrix along the branch cuts is $J$. 

It follows  from the local analysis of the Airy equation \eqref{Airy} that the relationship between bases in the regions of the two sides of a critical trajectory is given by the matrix $L$ as indicated in Fig. \ref{faces}.

 \begin{proposition}
 The jump matrices on the edges of the graph $\Sigma_Q$ are diagonal. Namely, suppose that two faces, 
 ${\mathscr D}$ and  $\tilde{\mathscr D}$, are separated by an edge $e$. Then 
 \be
 \varphi_\pm^{(\tilde{\mathscr D})}= e^{\pm\zeta_e} \varphi_\pm^{({\mathscr D})}
 \la{FGco}
 \ee
 where $\zeta_e\in \C/\sim$ is a monodromy parameter, and $\sim$ is the equivalence relation between $\zeta$ and $\zeta+\pi i$.
 \end{proposition}
 {\it Proof.}
For reference, consider the regions $\scr D= \scr D_a, \ \wt {\scr D}= \scr D_d$ separated by the edge $e=(z_1,z_2)$  in Fig. \ref{faces}.
It suffices to show that the analytic continuation of the recessive solutions $\varphi_-^{(\scr D_a)}$ is recessive also in the region $\scr D_d$.
This follows from the fact that the  real parts of  both flat coordinates $\int_{x_1}^x v$ and $\int_{x_2}^x v$ are necessarily positive  as we approach the pole $z_2$, as it follows from a local analysis. Thus on some segment of the edge $e= (z_1,z_2)$ of $\Sigma_Q$ near $z_2$, both flat coordinates have the same sign of the real part. Thus both solutions are recessive in a common domain and they must be proportional to each other. 

% 
%  Let us prove that two  distinguished bases of solutions are related by a diagonal jump matrix 
% when moving across the edge $e$ of $ \Sigma_Q$.  
%
%Now, consider the two neighbouring regions which  are denoted by $\scr D_a, \scr D_d$ in Fig.\ref{faces}. On the separating boundary the real parts of  $\int_{x_1}^x v$ and $\int_{x_2}^x v$ change sign and hence $\varphi_+^{{\scr D}_a}$ is dominant/recessive in overlapping regions of dominance/recessiveness (respectively) of $\varphi_+^{{\scr D}_d}$. Therefore they must be proportional by an $x$--independent constant (which, of course, depends on $\hbar$).  Similar reasoning works  for the pair $\varphi_-^{{\scr D}_a}, \varphi_+^{{\scr D}_d}$ and for the regions $\scr D_{c, b}$:
%\be
%\varphi_\pm ^{{\scr D}_a}(x;\hbar)=e^{\pm \zeta_e(\hbar)} \varphi_\pm ^{{\scr D}_d}(x;\hbar)
%\la{phiphi}
%\ee
Now it follows from the definition of $\varphi_+^{(\scr D_a)}$ and $\varphi_+^{(\scr D_d)}$ that they  are also proportional to each other. This can be seen  by repeating the same argument as above near $z_1$ where their analytic continuation coincides with  $\varphi_-^{(\scr D_b)}$ and   $\varphi_-^{(\scr D_c)}$, respectively,  by Def. \ref{def2}. 
 Therefore, the jumps on the edges of $\Sigma_Q$ are diagonal.
 \QED

% we have $\varphi_\pm^{(a)} \simeq f_\pm^{(1)}$ and $\varphi_\pm^{(d)} \simeq f_\pm^{(2)}$. The ratio of the formal solutions $f_+^{(1)}/f_+^{(2)}$ is a formal series  constant in $z$ given precisely by the exponential of the Voros symbol. Since the asymptotic expansion of the product of two functions (of $\hbar$) is the product of the corresponding asymptotic expansions, we 
The numbers $\zeta_e$ from (\ref{FGco}) are nothing but the complex shear coordinates on the $PSL(2)$ character variety (the monodromy manifold).

Let us now establish the link between the $\hbar$-expansion of the numbers $\zeta_e$ and the integrals of WKB differentials.
Consider for example the regions $\scr D_{a,d}$ of the shaded  quadrilateral with vertices $(x_1,z_1,x_2,z_2)$.
The first set $(\varphi^{{\scr D}_a}_+,\varphi^{{\scr D}_a}_-)$ has the asymptotic expansion given by formal solutions $f^{(1)}_\pm$
normalized at $x_1$:
\be
\varphi_\pm^{{\scr D}_a}\sim f^{(1)}_\pm
\la{as1}
\ee
The second set $(\varphi^{{\scr D}_d}_+,\varphi^{{\scr D}_d}_-)$ has the asymptotic expansion given by formal solutions $f^{(2)}_\pm$
normalized at $x_2$:
\be
\varphi_\pm^{{\scr D}_d}\sim f^{(2)}_\pm
\la{as2}
\ee
Since (in the sense of formal power series in $\hbar$)
\be
\int_{x_1}^x s_{odd}v= \int_{x_1}^{x_2} s_{odd}v+ \int_{x_2}^x s_{odd}v
\ee
the formal solutions  normalized at $x_1$ and $x_2$ can be related by the following formula which should be interpreted in the sense of formal power series in $\hbar$:
\be
f_\pm^{(1)}= f_\pm ^{(2)}\exp\left[  \pm \int_{x_1}^{x_2} s_{odd}v  \right]
\la{formalrel}
\ee
or, equivalently,
\be
f_\pm^{(1)}= f_\pm ^{(2)}\exp\left[ \pm\f{1}{2}  \int_{\ell_e} s_{odd}v  \right]\;.
\la{formalrel1}
\ee
Since the asymptotic expansion of a product is the product of the asymptotic expansions, we can assert the following proposition.

\begin{proposition}\la{asFG}
The formal asymptotic expansion of the entry $\zeta_e$ of the jump matrices on the edge $e$ of $\Sigma_Q$ 
is expressed in terms of periods of $s_{odd}v$ over cycle $\ell_e$ (Fig.\ref{faces}) as follows
\be
\zeta_e \sim \f{1}{2}\int_{\ell_e} v_{odd}
\la{zetas}
\ee
with $v_{odd}=s_{odd}v$.
The relation (\ref{zetas}) is understood in $PSL(2)$ sense i.e. up to an addition of   $
\pi i k$ for $k\in\Z$.

In terms of homological shear coordinates $\FG_\ell$ defined by (\ref{ile}) for each $\ell\in H_-$ the 
expansion (\ref{zetas}) can be written as follows:
be
\be
\FG_\ell\sim \int_{\ell} v_{odd}=\f{1}{\hbar} \int_{\ell} v+\hbar \int_{\ell} v_1+\hbar^3\int_{\ell} v_3 +\dots
\la{expFG}
\ee
%
%\la{WKBmain}
%\blue{NEED HERE??? Equivalently, the homological shear coordinates $\FG_\ell$ for any $\ell\in H_-$ have the following expansion in $\hbar$:
%\be
%\FG_\ell\sim \int_{\ell} v_{odd}=\f{1}{\hbar} \int_{\ell} v+\hbar \int_{\ell} v_1+\hbar^3\int_{\ell} v_3 +\dots
%\la{expFG}
%\ee

\end{proposition}

This proposition  is an analog of the theorem proved in \cite{Alleg}
where the reference projective connection $S_0$ in the equation (\ref{Sint}) is assumed to have 
poles of second order at the poles of $Q$ and biresidues $1/4$.

\begin{figure}[h]
\begin{center}
\begin{tikzpicture}[scale=1.0]
\begin{scope}[xshift=-57,rotate=60]
\Triangleplain{0}{0};
\node [below right]at (0.1,0) {$x_1$};
\node [below]at (120:2) {$z_4$};
\end{scope}
\begin{scope}[rotate=-00]
\Triangleplain{0}{0};
\node [below left]at (-0.1,0) {$x_2$};
\node [above]at (120:2) {$z_2$};
\node [below]at (-120:2) {$z_1$};
\node [below]at (-0:2) {$z_3$};
\end{scope}
\end{tikzpicture}
\end{center}
\caption{In $\mathbb P SL_2$  the edges carry the jump matrices indicated in Definition \ref{defJL}: the branch cuts carry the jump matrix $\pm J$; the blue edges carry the jump matrix $L$ and the red edges carry the diagonal jump matrices $Z_e$ depending on parameters.    Multiplying the fundamental solution by $\pm J$ in the shaded areas   gives a new Riemann--Hilbert problem where the resulting jumps are as indicated in Fig. \ref{stdjumps}. }
\label{chequered}
\end{figure}
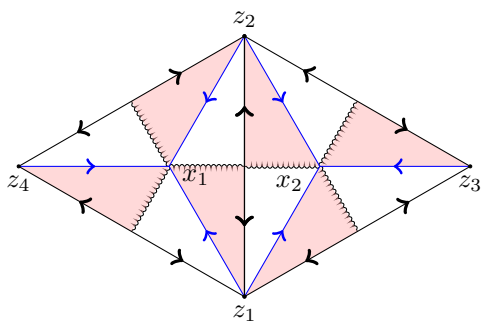

\vskip0.3cm
Finally, we describe the relationship of the Riemann-Hilbert problem \ref{RHpro} to the standard Riemann-Hilbert problem arising in shear coordinate parametrization of a $PSL(2)$ character variety. Such transformation amounts to 
multiplying by $\pm J$ in each region on the right of each critical trajectory, in the ``chequered'' pattern indicated schematically in Fig. \ref{chequered}.

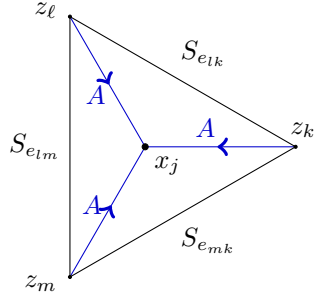
\begin{figure}[h]
\begin{center}
\begin{tikzpicture}[scale =1]
\begin{scope}
\draw [blue!80!black, postaction={decorate,decoration={markings,mark=at position 0.55 with {\arrow[line width=1.5pt]{<}}}}] (0,0) to 
node [pos=0.4, above] {$A$}
 (0:2);
\draw [blue!80!black, postaction={decorate,decoration={markings,mark=at position 0.55 with {\arrow[line width=1.5pt]{<}}}}] (0,0) to 
node [pos=0.4, left] {$A$}(120:2);
\draw [blue!80!black, 
postaction={decorate,decoration={markings,mark=at position 0.55 with {\arrow[line width=1.5pt]{>}}}}] (-120:2) to node [pos=0.55,left] {$A$} (0,0);

\draw  (0:2) to node[pos=0.55, below right] {$S_{e_{mk}}$}
(-120:2);
\draw  (0:2) to node[pos=0.55, above right] {$ S_{e_{lk}}$} (120:2);
\draw  (-120:2) to  node [pos=0.5, left] {$  S_{e_{lm}}$} (120:2);
\draw [fill](0:2) circle[radius=0.02];
\draw [fill](120:2) circle[radius=0.02];
%\node at (120:2) {$+$};
\draw [fill](-120:2) circle[radius=0.02];
%\node  at (-120:2) {$+$};
\node [left] at (120:2.1) {$z_\ell$};
\node [above] at (0:2.1) {$z_k$};
\node [left] at (-120:2.1) {$z_m$};
\draw [fill] node [below right] {$x_j$} circle[radius =0.04];
\end{scope}
\end{tikzpicture}

\end{center}
\caption{
The Riemann--Hilbert problem described in  Fig. \ref{faces}  is transformed to the standard 
Riemann-Hilbert problem shown here after multiplying   the vector of fundamental solutions within each shaded Stokes' region shown in Fig.\ref{chequered} by $J$.
}
 \label{stdjumps}
\end{figure}

As a result we get the Riemann-Hilbert problem on
the graph shown in Fig.\ref{stdjumps} where the matrix $A$ is given by (\ref{defA}) and
\be
S_e%:=-J \alpha^{\s_3} 
= \le({0 \ \  -e^{-\zeta_e} \atop 
 \!\!\!\!\!\!\!\!\! e^{\zeta_e} \ \ \ \,\, 0}\ri)
\ee
Note  that $S_e^{-1} = -S_e$, so that edges of the graph $\Sigma_Q$ are in fact un-oriented. 
%%%%%%%%%%%%%%%%%%%%%%%%%%%%%%%%%%%%%%%%%%%%%%%%%%%%%%
%\vskip3.0cm

\subsubsection{Toy example: WKB expansion of monodromy eigenvalues}

The Proposition \ref{asFG} has a simple special case, corresponding to 
$$
\ell=t_j^-=\f{1}{2}(t_j-t_j^\mu)
$$
where $t_j$ is the small positively-oriented loop around $z_j^{(1)}$. Then,  near $z_j$ we have
$Q=r_j^2({\d \xi}/{\xi})^2+\dots$ such that 
$r_j={\rm res}|_{z_j^{(1)}}v$
where $\xi$ is a local coordinate on $\CC$ near $z_j$ (and on $\Ch$ near $z_j^{(1)}$ and $z_j^{(2)}$).
Therefore,
$
\int_{t_j^-} v= 2\pi i  r_j
$.
On the other hand, the monodromy matrix $M_j$ has eigenvalues $m_j$, $m_j^{-1}$ and
$$
\FG_{t_j^-}=\log m_j
$$
(modulo an addition of an integer multiple of $\pi i$);
 the eigenvalues  are related to $r_j$ via (\ref{ra}):
 \be
 \f{r_j^2}{\hbar^2}=\f{\log m_j}{2\pi i}\left(\f{\log m_j}{2\pi i}-1\right)
 \la{vzeta}
 \ee
 and
 \be
 \f{\log m_j}{2\pi i}=\f{1}{2}\pm \f{r_j}{\hbar}\left(1+\f{\hbar^2}{4 r_j^2}\right)^{1/2}\;.
 \ee
% \red{Strange: 1/2 should not be there. Sign of $m_j$ in def of $\zeta$?}
 Using the  Taylor series
 $$
 (1+x)^{1/2}=1-\sum_{k=0}^\infty \f{2}{k+1} \left(\ba{c} {2k}\\k \ea\right)\left(-\f{x}{4}\right)^{k+1}\;
 $$
 the relation (\ref{vzeta}) gives the simplest example of WKB expansion of the generalized shear coordinates
 %(choose plus for $m_j$ and minus for $m_j^{-1}$ ??????):
 \be
 \f{\log m_j}{2\pi i}= \f{r_j}{\hbar}+ \f{1}{2}+\sum_{k=0}^\infty \f{1}{8(k+1)} \left(\ba{c} {2k}\\k \ea\right)\left(\f{\hbar}{ r_j}\right)^{2k+1}\;.
 \ee 
Therefore, we get the following WKB expansion of the homological shear coordinate  $\FG_{t_j^-}$ 
(as before, $1/2$ disappears since this relation is understood modulo an integer multiple of $\pi i$):
$$
\FG_{t_j^-}=\f{2\pi i r_j}{\hbar}+ 2\pi i \sum_{k=0}^\infty \f{1}{8(k+1)} \left(\ba{c} {2k}\\k \ea\right)\left(\f{\hbar}{ r_j}\right)^{2k+1}
$$
 %\red{minus lost somewhere, maybe $\zeta=-\log m$??}

Proposition \ref{asFG} implies the following classical formula \cite{KT05}.
\begin{corollary}
The residues of differentials $s_{2k+1}v$  at the poles $z_k$ are given by
\be
{\rm res}\big|_{z_k}v_{2k+1}= \f{1}{8(k+1)r_j^{2k+1}}\left(\ba{c} {2k}\\k \ea\right)
\ee
for $k=0,1,\dots$.
\end{corollary}

\subsection{WKB expansion of the generating function  and Bergman tau-function}

Here we discuss the WKB expansion of the generating function $\Gcal(\hbar)$  (\ref{Gdef}) for the equation 
(\ref{eqhbar}). This generating function is defined by the equation 
\be
\d \Gcal(\hbar)=\Fcal^* \theta_G(\hbar)- \theta_{hom}(\hbar)
\la{Ghbar}\ee
where the symplectic potentials $\theta_G$ and $\theta_{hom}$ are defined by (\ref{thG}) and (\ref{hompot}):
\be
\theta_{G}(\hbar)= \sum_{j=1}^{g_-} (\FG_{b_j^-} \d \FG_{a_j^-}- \FG_{a_j^-}  \d \FG_{b_j^-})
\la{spGh}
\ee
where $\FG_\ell(\hbar)$ is the homological shear coordinate corresponding to a loop $\ell\in H_-$ 
and
\be
\theta_{hom}=\f{1}{\hbar^2}\sum_{j=1}^{g_-} \left(B_k \d A_k- A_k\d B_k\right)
\la{thh}
\ee
where $(A_k=\int_{a_k^-}v\,,\,B_k=\int_{b_k^-}v)$ are period coordinates on $\Qcal_{g,n}[{\bf r}]$.

Proposition \ref{asFG} implies that      
for any contour $\ell\in H_-$ the homological shear coordinate $\FG_\ell$ on the $PSL(2)$ character variety has the following asymptotic expansion 
in powers of $\e$ as $\hbar\to 0$ (with slight abuse of notation our conventions imply $2\zeta_e=\FG_{\ell_e}$):
%\red{option: rename homological FG to $\rho$'s})
\be
\FG_\ell(\hbar)\sim   \f{1}{\hbar}  \int_{\ell} v +\hbar \int_{\ell} v_1+ \hbar^3 \int_{\ell} v_3 +\dots 
\la{aszeta}
\ee

Denote the periods of the differentials $v_{2k+1}=s_{2k+1}v$ by % \red{what about $\sqrt{2}$ here???}
\be
A_j^{(2k+1)}=\int_{a_j^-}v_{2k+1}\;, \hskip0.7cm
B_j^{(2k+1)}=\int_{b_j^-}v_{2k+1}
\ee
with $A_j=A_j^{(-1)}$ and $B_j=B_j^{(-1)}$.

According to (\ref{aszeta}) we have
\be
\FG_{a_j^-}= \f{1}{\hbar} A_j +\hbar A_j^{(1)}+O(\hbar^3) \;,\hskip0.7cm
\FG_{b_j^-}= \f{1}{\hbar} B_j +\hbar B_j^{(1)}+O(\hbar^3) 
\la{zz}
\ee

By plugging  (\ref{zz}) in (\ref{Ghbar}) we see that   the coefficient in front of $\hbar^{-2}$ in the expansion of  (\ref{Ghbar})   vanishes and
$$
\Fcal^*\theta_{G}(\hbar)-\theta_{hom}(\hbar)
$$
$$
=\sum_{j=1}^{g_-}  \sum_{r=-1}^\infty \hbar^{2r+2}\sum_{l+k=r}% \left[
\left(B^{(2l+1)}_j \d A^{(2k+1)}_j-A^{(2l+1)}_j\d B^{(2k+1)}_j \right)
%+\left(B^{(2k+1)}_j \d A^{(2l+1)}_j-A^{(2k+1)}_j \d B^{(2l+1)}_j\right)\right]
$$
$$
= \sum_{j=1}^{g_-}  \left[\left(     B_j \d A^{(1)}_j   -A_j \d B^{(1)}_j\right) +\left( B^{(1)}_j \d A_j - A^{(1)}_j \d B_j\right)\right] +\mathcal O(\hbar^2)
$$
\be
=\d\; \sum_{j=1}^{g_-}  \left(B_j A^{(1)}_j- A_j  B^{(1)}_j \right) +2\sum_{j=1}^{g_-} \left( B^{(1)}_j\d A_j-A^{(1)}_j \d B_j \right)+\mathcal O(\hbar^2)\;.
\la{thethe}
\ee

The  second sum in  (\ref{thethe})  is, up to a multiplicative constant, the differential of $\log \tau_B$ on $\Qcal_{g,n}[{\bf r}]$. Indeed  the equations for the tau-function (\ref{defBerg}) can be written as 
\be
12\pi i\, \d \log \tau_B=\sum_{j=1}^{g_-} \left(   B_j^{(1)}\d A_j- A_j^{(1)} \d B_{j}\right)
\ee
where $A_j^{(1)}$ and $B_j^{(1)}$ are periods of 
$$
v_1=s_1 v=\f{\Scal_B-\Scal_v}{2v}
$$

Consider   set of generators (\ref{homba}) of $H_-$ .
For any two abelian differentials $v,w$ on $\Ch$ with poles (possibly with residues) at $z_1,\dots,z_n$ 
 we introduce the pairing
\be
\langle v, w\rangle =\sum_{j=1}^{g_-}\left[\int_{b_j^-} v\;\int_{a_j^-} w-
\int_{a_j^-} v\;\int_{b_j^-} w\right]\;.
\la{pairing}
\ee
The pairing (\ref{pairing}) is invariant under  a change of generators (\ref{homba}) which is symplectic 
in the $(a_j^-,b_j^-)$ subspace and does not involve $t_j^-$.

%To formulate the next theorem which is the main result of this paper we start from 

Then (\ref{thethe}) implies 

\begin{theorem}\la{mainth}
 Let $\CC$ be a Riemann surface of genus $g$. Let $Q$ be a quadratic diffefrential on $\CC$
 with $n$ second order poles at $z_1,\dots,z_n$ and biresidues $r_1^2,\dots, r_n^2$.
 
 Consider the triangulation 
$\Sigma_Q$ (with vertices at $z_j$'s) and  the dual three-valent graph $\Sigma_Q^*$ 
(with vertices at the zeros $x_j$ of $Q$)
defined in App. \ref{SSQ}  and let 
$G$ be a spanning tree of 
 $\Sigma_Q$.
 Choose a contour $\ell$ connecting  $z_1$ with one of the vertices (denoted by $x_1$) of the corresponding triangle and introduce the tree graph $\Gh=\pi^{-1}(G\cup \ell)$. Choose 
the contours representing canonical cycles on $\Ch$ such that $\Gh$ lies entirely in the corresponding fundamental polygon of $\Ch$. Using these data (which in particular define the Torelli marking of $\CC$) we define the distinguished local coordinates near poles $z_j$ (\ref{xijint}), the Bergman tau-function 
$\tau_B$ (\ref{taupint}), the Bergman projective connection $S_B$, the homological coordinates 
$(A_j,B_j)_{j=1}^{g_-}$ (\ref{homco}) and the homological shear coordinates $\FG_{a_j^-}$, $\FG_{b_j^-}$ (\ref{shearperiods}).
 
  Consider the 
 differential equation 
\be
\phi''+\left(\f{1}{2} S_B - \f{1}{\hbar^2} Q\right)\phi=0
\la{eqhbar1}
\ee
on $\CC$.
%where $S_B$ is the Bergman projective connection corresponding to some choice of canonical basis 
%of cycles on $\CC$. 
Denote by $\Fcal$ the monodromy map between the moduli space $\Qcal_{g,n}[{\bf r}/\hbar]$
and the symplectic leaf  $CV_{g,n}[{\bf m}({\hbar})]$ of the $PSL(2)$ character variety, where each $m_j(\hbar)$ is expressed via $r_j$  by (\ref{vzeta}).

%\red{need triangulation, then tree, then definition of odd part of homologies..}

%Choose a set of generators (\ref{homba}) in the odd part $H_-$ of the homology group $\pi_1(\Ch,\R)$ where $\Ch$ is the canonical cover $v^2=Q$ and 

Introduce the symplectic potential $\theta_{hom}$ (\ref{thh})
of the homological symplectic form on $\Qcal_{g,n}[{\bf r}/\hbar]$ and symplectic potential $\theta_G$ (\ref{spGh}) for the Goldman symplectic form on $CV_{g,n}[{\bf m}(\hbar)]$.

The  generating function $\Gcal$  of the monodromy symplectomorphism between $\Qcal_{g,n}[{\bf r}/\hbar]$
and $CV_{g,n}[{\bf m}(\hbar)]$ is defined by
$$
 \d \Gcal(\hbar)=\Fcal^* \theta_G(\hbar)- \theta_{hom}(\hbar)
 $$
 
 Under these assumptions the function $\Gcal$  has the following asymptotics  as $\hbar\to 0$
\be
\Gcal= -12\pi i\, \log \tau_B +2 \langle v_{-1}, v_1\rangle + O(\hbar^2 )
\la{asYY}
\ee
where $v_{-1}=v$, $v_1=s_1 v$ are the first two non-vanishing terms  (\ref{defvk}) in the WKB expansion.
% $\tau_B$ is the Bergman tau-function (\ref{taupint}) defined using the same Torelli marking as the Bergman projective connection $S_B$.  \red{The choice of the  integration contours $l_j=[x_1,z_j]$ in the definition (\ref{xijint}) 
%of the distinguished local parameters used in  (\ref{taupint}) should agree with the choice of the 
%generators (\ref{sstar}) in $H_-$ such that $\kappa_j^-=\f{1}{2}(l_j-l_j^\mu)$.}
\end{theorem}

\section{Open problems}
\la{openpro}

Here we list a few  open problems related to this work

\begin{itemize}
\item
When the Torelli marking of the Riemann surface $\CC$ changes
by an $Sp(2g,\Z)$ matrix
$\left(\ba{cc} C & D \\ A & B\ea\right)$, the Bergman projective connection $S_B$ also changes; however, the monodromy map remains a symplectomorphism.  When $n=0$ (holomorphic case) or $r_j=0$ (the potentials with simple zeros)  the function $\Gcal$ is gaining an additional term of
$-12\pi i \log  {\rm det} (C\omega+D)$ \cite{BKN,Kor}. We expect this to hold true for $r_j\neq 0$, but 
don't have a complete proof at the moment. The proof of \cite{BKN,Kor} is based on the identification of the canonical and homological symplectic structures for moduli spaces $\Mcal_g$ and $\Mcal_{g,n}$.

\item
The symplectic potential $\theta_G$ for the Goldman bracket depends on the choice of the 
triangulation $\Sigma$ used to define the homological shear coordinates. If $\Sigma$ changes the
potential $\theta_G$ changes by a combination of Roger's dilogarithms, although in the current setting
this change is not known explicitly (although it is known for an alternative choice of $\theta_G$, used in 
\cite{BK2}).

%Transformation of $\Gcal$ under the change of triangulation and WKB expansion of the dilogarithm 
%multiplier.
\item
We expect that the monodromy symplectomorphism can be extended to the full character variety by introducing variables conjugate to the residues $r_j$, similarly to \cite{BK2}, where this scheme was realized for fuchsian systems on a Riemann sphere. 
Study of the corresponding generating function and its WKB expansion would be an important
problem.

\item
To map the space $\Qcal_{g,n}[{\bf r}]$ to the cotangent bundle to $\Mcal_{g,n}$ and then proceed as in    \cite{BKN,Kor}one can potentially follow the following scheme: instead of $S_B$ one can try to use a singular projective connection with 
poles at $z_j$ and biresidues $r_j^2$. In this way one could split the potential in the equation (\ref{eqhbar1}) 
in a different way, and assume that $Q$ has only first order poles at $z_j$, and therefore can be identified with cotangent vector to $\Mcal_{g,n}$. The problem here is that it's unclear how to make such reference projective connection depend only on a point of $\Mcal_{g,n}$ (plus some discrete data, like Torelli marking). At the moment we are not aware of a good candidate to such projective connection.

\item
The remaining terms $G_1,\;G_2,\dots $  of the asymptotic expansion  in (\ref{asYY}) satisfy a system of equations involving periods of the other differentials $v_{2k+1}$; an explicit evaluation of these coefficients 
in terms of theta-functions presents an interesting problem; these coefficients are expected to be related  to 
constructions arising in the theory of topological recursion \cite{EO,CEO}.
%Recently an expression for $G_1$ was derived in \cite{Klimov}.

\item
The WKB formalizm can be modified by adding a term of the form $\f{1}{\hbar} Q_1$ to the potential 
of equation (\ref{eqhbar1}); then the canonical cover become $\hbar$-dependent. Such approach was taken in \cite{Bri}. Modification of our framework to this case is an important open problem.

\item
We expect that the theorem that $\Fcal$ is a symplectomorphism admits a simpler proof than the one proposed in \cite{BKN} and adopted here to the case of second order. Such proof should follow the computation of \cite{BK2iso} adopted to the higher genus situation; the alternative  proof should also be useful 
in proving the symplectic nature of the extended monodromy map.  

\end{itemize}

{\bf Acknowledgements.}
We thank D. Allegretti, T. Bridgeland, A. Nietzke and C. Norton for illuminating discussions. 
The work of M.B. was supported in part by the Natural Sciences and Engineering Research Council of Canada (NSERC) grant RGPIN-2016-06660.
The work of D.K. was supported in part by the NSERC grant
RGPIN/3827-2015.
This  work was also supported by the National Science Foundation under Grant No. DMS-1440140 while the authors were in residence at the Mathematical Sciences Research Institute in Berkeley, California, during the Fall 2019 semester
{\it Holomorphic Differentials in Mathematics and Physics}.

\appendix
\section{Topological double covers and Darboux coordinates for $PSL(2)$ Goldman bracket} 
\la{monre}

For an $n$-punctured Riemann surface $\CC$ consider the space of $PSL(2)$ monodromy representation of $\pi_1(C\setminus\{z_j\}_{j=1}^n)$ (the character variety). The $PSL(2)$ character variety is equipped with Goldman's bracket \cite{Gold84}

%\subsection{Goldman bracket in complex shear coordinates}

%\subsection{Fock-Goncharov (shear) coordinates on character variety}

\be
\{{\rm tr}M_\g,\;{\rm tr}M_\gt\}_G=\f{1}{2}\sum_{p\in \g\circ\gt}\nu(p)\left( {\rm tr}M_{\g_p \gt}- {\rm tr}M_{\g_p \gt^{-1}}\right)\;.
\la{Goldapp}
\ee

The Goldman bracket (\ref{Goldint}) is degenerate; the Casimirs are
the eigenvalues $m_j$ of the monodromy matrices.

The symplectic form which inverts the Goldman bracket on a symplectic leaf $m_j=const$ can be written in terms of logarithmic complex shear coordinates as follows.
Consider a graph $\Sigma$  with $n$ vertices at poles $z_1,\dots,z_n$ of $Q$ which defines a  triangulation  of the surface $\CC$. 
Denote the edges by $e_1,\dots,e_{6g-6+3n}$ and faces by $f_1,\dots,f_{4g-4+2n}$.

To each edge $e$ of the graph  $\Sigma$  we assign a  coordinate $\zeta_e\in \C$ which (up to a factor $1/2$)  is the logarithm of the corresponding ``complex shear"   coordinate (the simplest example of the general 
Fock-Goncharov coordinate; in the real-analytic case the shear coordinates were proposed by Thurston, see \cite{Thurston}).

The monodromy matrices are expressed in terms of the edge coordinates as discussed in \cite{GMN},
see also \cite{BK2} for the description close to the framework of this paper.

The Casimirs $m_j$ of the Goldman's bracket (\ref{Goldint})    are expressed in terms of coordinates $\zeta_e$ as follows (since we are dealing with the $PSL(2)$ representation this relation  is understood up to a sign):
\be
m_j=\exp\left(\sum_{e\perp z_j} \zeta_e\right)\;,\hskip0.8cm j=1,\dots,n
\la{mj}
\ee
where the notation $e\perp z_j$ means that $e$ is one of the edges of $\Sigma$ connected to $z_j$.
Consider a ciliation  of  the graph $\Sigma$ i.e. assume that  the edges $e$ 
attached to a vertex $z_j$ 
are ordered counter-clockwise at each vertex \cite{BK2} starting from the "cilium". Then the following proposition holds (see Th.4.1 and (7.5) of \cite{BK2} and also \cite{Chekhov}): 
%\red{monodromies in terms of $\zeta_e$??}
%\todo[inline]{Sign and factors wrong}
\begin{proposition}
The two-form 
\be
\Omega_{G}=\sum_{v\in V(\Sigma)} \sum_{e,e'\perp v,\;e'\prec e} \d \zeta_{e'} \wedge \d \zeta_{e}
\la{Omegamon}
\ee
%\red{need to look at this factor of 2 - it looks like it should not be there..}
is the inverse of the Poisson bracket (\ref{Goldint}) on each symplectic leaf $m_j=const$. The ordering 
$\prec$ of edges attached to a given vertex is determined by the choice of the ciliation; the form (\ref{Omegamon}) is independent of ciliation on a symplectic leaf.
\end{proposition}

%Notice that the coefficients of the form (\ref{Omegamon}) are constant; inversion of this expression on the
%symplectic leaf allows to obtain the expression for the associated Poisson tensor (Corollary 7.1 of \cite{BK2}):

In turn, the inversion of the form (\ref{Omegamon}) is given by the following proposition
(see \cite{Fock1,Chekhov07} andTh.7.2 of \cite{BK2}):
\begin{proposition}
The Goldman Poisson tensor,  which is an inverse of the symplectic form (\ref{Omegamon}) on a symplectic leaf $m_j=const$,   is given by %\blue{wrong sign??}
\be
{\bf P}_G= -\f{1}{4}\sum_{f\in F(\Sigma)}
\left(\f{\p}{\p \zeta_{e_1}}\wedge  \f{\p}{\p \zeta_{e_2}}+\f{\p}{\p \zeta_{e_2}}\wedge  \f{\p}{\p \zeta_{e_3}}
+\f{\p}{\p \zeta_{e_3}}\wedge  \f{\p}{\p \zeta_{e_1}}\right)
\la{PTen}
\ee
 where $e_1$ $e_2$ and $e_3$ are the edges (counted counter-clockwise) which form the boundary of the face $f$ of $\Sigma$. %\red{Was it also written in old Fock??}.
\end{proposition}

\begin{figure}[htb]
\begin{center}
\includegraphics[width=0.3\textwidth]{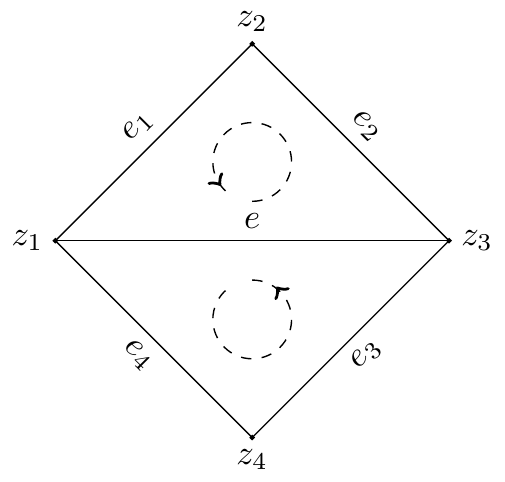}
\end{center}
\caption{Ordering of edges of  faces of $\Sigma_Q$.}
\label{two_tri}
\end{figure}

The form of the Poisson tensor (\ref{PTen}) is equivalent to the statement that the variable $\zeta$, associated to an edge $e$, has non-vanishing Poisson brackets only with four edges forming the boundary of two faces adjacent to $e$. Namely, in the notations shown in Figure \ref{two_tri} we have 
\be
\{\zeta,\zeta_2\}_G=-\f{1}{4}\;,\hskip0.5cm
\{\zeta,\zeta_1\}_G=\f{1}{4}\;,\hskip0.5cm
\{\zeta,\zeta_4\}_G=-\f{1}{4}\;,\hskip0.5cm
\{\zeta,\zeta_3\}_G=\f{1}{4}\;.
\la{PBzeta}
\ee
%For the reader $dp\wedge dq$, $\{p,q\}=1$, $P=\p_p\wedge\p_q$}
%We notice that the normalization factor in Th.7.1 of \cite{BK2}) is different 

\subsection{Double covers $\Ch_{\Sigma}$}   
\la{monre1}

According to expressions for the Goldman's Poisson tensor (\ref{PTen}) and the corresponding 
symplectic form (\ref{Omegamon})  the coordinates $\zeta_j$ are ``almost Darboux" - their Poisson brackets are 
constant. To describe their linear combinations which give the actual Darboux coordinates one can use the
following topological construction. % this construction does not use the complex structure of $\CC$.
 For each triangle $f_j$,  of the triangulation $\Sigma$ we fix a point $x_j$ in the interior of $f_j$
 (here $j=1,\dots,N$ and $N=4g-4+2n$).

%Consider  a point $x_j$  inside of each face $f_j$ of the triangulation $\Sigma$ for each $j=1,\dots, N$ 
%for 
%$$
%N=4g-4+2n
%$$
Consider the tri-valent graph 
$\Sigma^*$ dual to $\Sigma$ with vertices at the points $x_j$.  We denote by $e^*$ the edge of $\Sigma^*$ which is dual to the edge $e$ of $\Sigma$.

%The number $N$ of vertices of $\Sigma^*$ is even. Consider some perfect matching $p$ of $\Sigma^*$
%(a perfect matching means that  the total set of vertices of $\Sigma^*$ is split into non-overlapping pairs of vertices connected by an edge, see \cite{Kenyon}).

\begin{definition}
%For a given perfect matching $p$ of the graph $\Sigma^*$ 
The double cover $\Ch_{\Sigma}$ is defined  by gluing two copies of $\CC$ 
via branch cuts chosen along {\bf all edges} of $\Sigma^*$. 
%which belong to the bi-partition $p$. 

\end{definition}

According to this definition, all points $x_j$ are  the branch points of $\Ch_{\Sigma}$ with three branch cuts meeting at each $x_j$.

\begin{figure}[htb]
\begin{center}
\includegraphics[width=5cm]{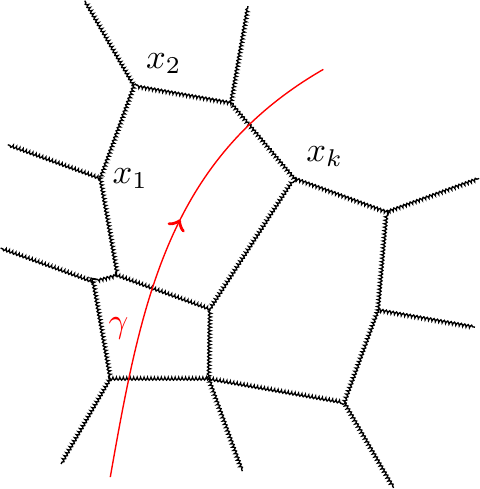}
\end{center}
\caption{A contour $\gamma\in \pi_1(\CC\setminus\{x_j\})$ crossing the branch cuts i.e. all edges of $\Sigma_Q^*$}
\label{gammaint}
\end{figure}

Topologically, the branched double cover is uniquely defined by a homomorphism $h$ of the 
fundamental group $\pi_1(\CC\setminus \{x_i\}_{i=1}^N, x_0)$ (for some initial point $x_0\in \CC$ which does not coincide with any branch point $x_j$) into the symmetric group
$S_2$. 
This homomorphism 
%for the covering $\Ch_{\Sigma}$ 
is constructed as follows.
Denote the standard generators of $\pi_1(\CC\setminus \{x_i\}_{i=1}^N, x_0)$ by $\{\alpha_j,\beta_j\}_{j=1}^g$ and $\{\delta_k\}_{k=1}^N$; these generators satisfy the relation
\be
\delta_N\dots\delta_1\prod_{j=1}^g[\alpha_j, \beta_j]= id\;.
\la{fgrel}\ee
For any $\gamma\in \pi_1(\CC\setminus \{x_i\}_{i=1}^N, x_0)$ %which we assume not to pass through the vertices $x_j$ 
we define
\be
\label{rindex}
h(\gamma)= (1\;\; 2)^{r(\gamma,\Sigma)}
\ee
where $(1\;\;2)$ is the generator of $S_2$ and $r(\gamma,\Sigma)$ is the number of intersections of $\gamma$ with edges of 
$\Sigma^*$
%belonging to the bi-partition $p$ 
(Fig.\ref{gammaint}). Clearly $r$ is odd for each $\delta_k$; thus $h(\delta_k)=(1\;\; 2)$ and each $x_k$ is indeed a branch point of $\Ch_\Sigma$.

The genus of $\Ch_{\Sigma}$ equals to
\be
\gh=g+g_-
\la{ghdec}
\ee
where
\be
g_-=3g-3+n\;.
\la{gmin}
\ee

The projection $\Ch_{\Sigma} \to \CC$ is denoted by $\pi$ and the natural involution on $\Ch_\Sigma$ by $\mu$.

There are two points on $\Ch_{\Sigma}$ which project to each vertex $z_j$ of $\Sigma$; we denote 
them by 
$z_j^{(1)}$ and   $z_j^{(2)}$ % \red{correct above the notations} 
with
$$
z_j^{(2)}=(z_j^{(1)})^\mu\;.
$$

For  each (open) face $f_j^*$ of $\Sigma^*$ its lift $\pi^{-1}(f_j^*)$ to 
$\Ch_{\Sigma}$ consists of two disks, $f_j^{* \,(1)}$ and $f_j^{* \,(2)}$; we assume that $z_j^{(1)}\in f_j^{*\,(1)}$
and $z_j^{(2)}\in f_j^{*\,(2)}$.

Consider the homology group $H_1(\Ch_\Sigma\setminus \{z_j^{(1,2)}\}_{j=1}^n)$ and consider its decomposition into symmetric and skew-symmetric subspaces under the involution $\mu$:
$$
H_1(\Ch_\Sigma\setminus \{z_j^{(1)},\,z_j^{(2)}\}_{j=1}^n)=H_+\oplus H_-
$$
where $H_+$ can be identified with $H_1(\CC\setminus \{z_j\}_{j=1}^n)$; thus
${\rm dim} H_+=2g+n-1$. For the group $H_-$ we have
$$
\dim H_-=6g-6+3n=2 g_-+n
$$
where $g_-$ is given by (\ref{gmin}).

The rank of the intersection pairing on $H_-$ equals $6g-6+2n=2 g_-$.

%It may seem that the cover depends on the choice of triangulation $\Sigma$. The next proposition shows that this is 

The next proposition shows that the covering $\Ch_{\Sigma}$ does not depend of the choice of the triangulation
$\Sigma$ as long as the sets of $x_j$ and $z_j$ remain the same:

\begin{proposition}\label{propSigma}
Let $\Sigma$ and $\wt\Sigma$ be two triangulations of $\CC$ with the same vertices and the same points $x_j$ chosen within faces. Then the covers $\Ch_{\wt\Sigma}$
and  $\Ch_{\Sigma}$ are topologically (and holomorphically) equivalent.

%Let $\wt \Sigma$ be another triangulation with the same vertices. The cover $\Ch_{\wt\Sigma}$ is topologically (and holomorphically) equivalent to $\Ch_{\Sigma}$.
\end{proposition}
{\bf Proof.}
Let us  use the fact that any two triangulations, $\Sigma$ and $\wt\Sigma$, can be connected by a sequence of flips of diagonals of quadrilaterals formed by two neighbouring triangles. The corresponding dual graphs are
then connected by  a sequence of Whitehead moves. Thus it is  enough to show that the index $r(\gamma,\Sigma)$ in \eqref{rindex} remains of the same parity under an elementary 
Whitehead move  (recall that we assume that the points $x_j$ are the same for the two triangulations).

\begin{figure}[htb]
\begin{center}
\includegraphics[width=0.8\textwidth]{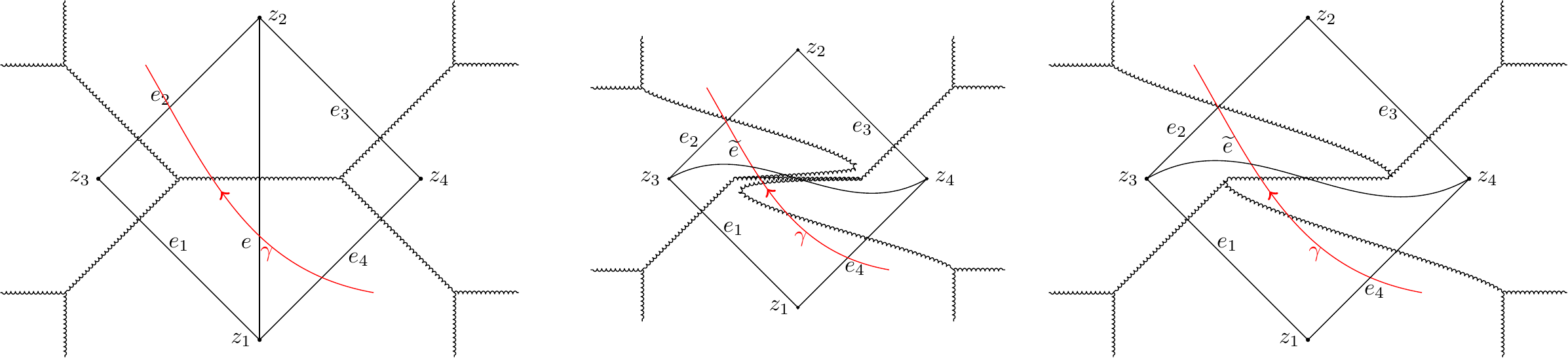}
\end{center}
\caption{Transformation of cuts after a Whitehead move. In the middle frame we show the isotopy of cuts between the two. }
\label{WH}
\end{figure}

The effect of the Whitehead move involving two triangles containing the points $x_1,x_2$ is equivalent to ``sliding'' two branch-cuts along the cut connecting $x_1,x_2$ as illustrated in Fig. \ref{WH}.  Clearly the indices $r(\gamma,\Sigma)$ and  $r(\gamma, \widehat \Sigma)$ have the same parity.
\QED

Consider two sets of generators of $H_-$.  The first set  is 
denoted by 
\be
\{s_j\} _{j=1}^{2g_-+n}= \{a_i^-,b_i^-\}_{i=1}^{g_-},\;\{t_i^-\}_{i=1}^n
\la{gen1}\ee
 with the intersection index 
\be
a_i^-\circ b_j^-=\f{\delta_{ij}}{2}\;,\hskip0.7cm t_i^-\circ s_j=0
%\la{intind}
\ee
for all $s_j\in H_-$, so that the intersection index of $t_i^-$ with all  generators equals to $0$. The generator $t_i^-$ 
is equal to $\f{1}{2}(t_i-t_i^\mu)$ with $t_i$ being a small  positively-oriented loop around $z_i^{(1)}$.

\begin{figure}[htb]
\begin{center}
\includegraphics[width=7cm]{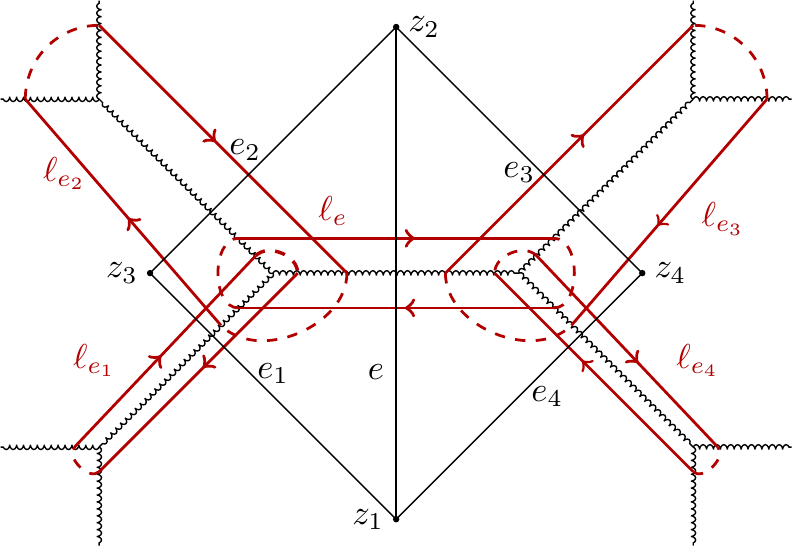}
\end{center}
\caption{Cycles $\ell_e$ and their intersection indices:  $ \ell_e \circ \ell_{e_1} = 1$,
$ \ell_e \circ \ell_{e_2} = -1$,
$ \ell_e \circ \ell_{e_3} = 1$,
$ \ell_e \circ \ell_{e_4} = -1$. }
\label{contl}
\end{figure}
%We denote by $z_j^{(1)}$ the poles lying on the "first" sheet of $\Ch^p_\Sigma$. 

We denote the second set of generators  by $\{\ell_{e_j}\}_{j=1}^{2g_-+n}$; their number  equals to the number of edges of $\Sigma$. The generator $\ell_e$ goes clockwise around the edge $e^*$ of $\Sigma^*$  with the orientation chosen  as shown 
in Fig. \ref{contl}.

\begin{figure}[htb]
\begin{center}
\includegraphics[width=4cm]{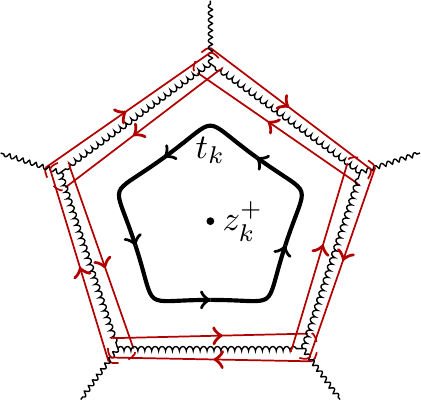}
\end{center}
\caption{Expressing the cycle $t_k^-=\f{1}{2}(t_k+t_k^\mu)$ via $l_e$ for $e\in \partial f_k$}
\label{TL}
\end{figure}

The cycles $t_k^-$ can be easily expressed in terms of the cycles $\ell_e$ corresponding to the face $f_k^*$, as can be seen from Fig.\ref{TL}:
\be
t_k^-=\f{1}{2}\sum_{e\in\p f_k} \ell_e\;.
\la{tl}
\ee

It is important to observe that each contour $\ell_e$ has non-vanishing intersection index with exactly four neighbours; these intersection indeces are seen in Fig. \ref{contl}:
\be
\ell_e\circ \ell_{e_1}=1\;,\hskip0.7cm 
\ell_e\circ \ell_{e_2}=-1\;,\hskip0.7cm 
\ell_e\circ \ell_{e_3}=1\;,\hskip0.7cm 
\ell_e\circ \ell_{e_4}=-1\;,\hskip0.7cm 
\la{intindl}
\ee

\subsection{Darboux coordinates for Goldman bracket: homological  shear coordinates}
\la{app2}

Comparing the intersection indices (\ref{intindl})  with the Poisson brackets (\ref{PBzeta}) we get the following lemma:
\begin{lemma}
The Goldman Poisson brackets (\ref{PBzeta}) between the coordinates $\zeta_{e}$ can be expressed via the intersection indices of the cycles $\ell_e\in H_-$ as follows:
\be
\{\zeta_e,\zeta_{\tilde{e}}\}_G=\f{1}{4} 
 \ell_e\circ  \ell_{\tilde{e}}   \;.
\la{PBint}
\ee 
\end{lemma}

%\red{Need differerent sign!}

This lemma allows us to extend by linearity the notion of the logarithmic complex shear coordinate to any cycle $\ell\in H_-$ expressed via linear combination of $\ell_{e_k}$.
\begin{definition}
The {\bf homological logarithmic shear coordinate} $\zeta_\ell$ corresponding to a cycle $\ell\in H_-$ given by
\be
\ell=\sum_{j=1}^{2 g_-+n } k_j \ell_{e_j}
%\la{ile}
\ee
is defined to be 
\be
\FG_\ell=2\sum_{j=1}^{2 g_-+n } k_j \zeta_{e_j}\;.
\la{ile}
\ee
\end{definition}

The Goldman Poisson brackets between  homological shear coordinates corresponding to any two cycles $\ell,\tilde{\ell}\in H_-$ then
follow from (\ref{PBint}):
\be
\{\FG_\ell,\FG_{\tilde{\ell}}\}_G=%\red{ \tilde{\ell}\circ \ell=-}\hskip0.5cm 
\ell\circ \tilde{\ell}  \;.
%\la{PBint}
\ee 

To construct the Darboux coordinates on the symplectic leaf of the Goldman bracket we express the 
cycles 
$(a_i^-,b_i^-)_{i=1}^{g_-}$ (which form a  part of the basis (\ref{gen1})) as linear combinations  of $\ell_{e_j}$:
\be
a_j^-=\sum_{k=1}^{2 g_-+n}\alpha_{jk} \ell_{e_k}\;;\hskip0.7cm
b_j^-=\sum_{k=1}^{2 g_-+n}\beta_{jk} \ell_{e_k}
\la{abl}
\ee
where $\alpha_{jk}\in \Z/2$.

Then, taking also (\ref{tl}) into account, we arrive at the following proposition:

\begin{proposition}
The following linear combinations of  shear coordinates:
\be
 \FG_{a_j^-} =2\sum_{k=1}^{2g_-+n}\alpha_{jk}\, \zeta_{e_k}\;;\hskip0.7cm
 \FG_{b_j^-}=2\sum_{k=1}^{2g_-+n}\beta_{jk} \, \zeta_{e_k}\;;\hskip0.7cm
  \FG_{t_k^-}=\sum_{e\in\p f_k} \, \zeta_{e}
\la{shearperiods}
\ee
have the Poisson brackets
\be
 \{\FG_{a_j^-},\;  \FG_{b_k^-}\}_G=\f{\delta_{jk}}{2}\;;\hskip0.7cm
  \{\FG_{a_j^-},\;  \FG_{a_k^-}\}_G= \{\FG_{b_j^-},\;  \FG_{b_k^-}\}_G=0
 \la{PBZab}
\ee
while  $\FG_{t_j^-}$ lie in the center of the Poisson algebra.
% they are equal to $\log m_j$ where the Casimirs $m_j$ are given by (\ref{mj}).
\end{proposition}

The coordinates $\FG_{t^-_j}$ are related to the monodromy eigenvalue as follows:
\be
\FG_{t_j^-}=\log m_j %\red{+\pi i ?}
\la{tm}
\ee
(as usual, in the $PSL(2)$ case this equality is understood moduli an integer  multiple of $\pi i$).

The corresponding symplectic form is therefore given by   the following proposition (our assumption in this paper is that the
Poisson structure $\{p,q\}=1$ corresponds to the symplectic form $\d p\wedge \d q$).

\begin{proposition}
The $PSL(2,\C)$ Goldman's symplectic form on a leaf $\{\FG_{t^-_j}=const\}_{j=1}^n$ is given by 
\be
\Omega_{G}=2 \sum_{j=1}^{g_-} \d \FG_{a_j^-} \wedge \d \FG_{b_j^-}\;.
\la{GolDar}
\ee
\end{proposition}

The form (\ref{GolDar}) is invariant if a cycle $a_j^-$ or $b_j^-$ is changed by a linear combination of the 
cycles $t_j^-$; under such transformation the respective coordinates $\FG_{a_j}^-$ and $\FG_{b_j}^-$
change by a linear combination of $\log m_k$ and the differentials  $\d \FG_{a_j^-}$  and  $\d \FG_{b_j^-}$
do not change on the symplectic leaf.

%However, to fix the arbitrariness % (for a chosen covering $\Ch_\Sigma^p$)
 %in the choice of symplectic potential for the form $\Omega_{G}$ we should be 
%more specific about the choice of  $\zeta_{a_j}^-$ and $\zeta_{b_j}^-$.

%\red{!!!!!!!!!!  Branch cuts!!!!}

%To fix the arbitrariness in the choice of coordinates $\zeta_{a_j^-}$ and $\zeta_{b_j^-} $  we fix the
%fundamental polygon $\CC_0$ for $\CC$ placing the vertex at the zero $x_1$ (Fig.\ref{funpol}).
%Then $\Ch_\Sigma$ can be represented as two-sheeted cover of $\CC_0$ with an appropriate identification of 
%sides of $\CC_0$.

%We choose the disc $D\subset \CC_0$ which contains all the poles $z_1,\dots,z_n$ and connect it by a contour $\kappa$ to the disc $D$. Lifting to $\Ch_\Sigma$ we get two discs, $D^{(1)}$ 
%(which contains $z_1^{(1)},\dots, z_n^{(1)}$) and $D^{(2)}$ 
%(which contains $z_1^{(2)},\dots, z_n^{(2)}$). These discs are connected by the contour $\kappa^{(1)}\cup\kappa^{(2)}$ which passes through $x_1$, see Fig.\ref{funpol}.

%Then, assuming that the canonical basis on $\Ch_\Sigma$ is chosen as shown in Fig.\ref{funpol} we
%define the cycles $\{a_j^-, b_j^-\}$ via (\ref{abm1}), (\ref{abm}). 

\begin{definition}
Let the homological shear coordinates 
$(\FG_{a_j^-},\FG_{b_j^-})$ be defined as discussed above.
Then the symplectic potential of the form $-\Omega_G$   on a symplectic leaf 
$$\{\FG_{t_i^-}={\rm const}\}_{i=1}^n$$ 
(or, equivalently, $ \{m_i={\rm const}\}_{i=1}^n$)  is defined by
\be
\theta_{G}= \sum_{j=1}^{g_-} (\FG_{b_j^-}  \d \FG_{a_j^-}-\FG_{a_j^-}  \d \FG_{b_j^-})
\la{potmon}
\ee
such that $\d \theta_{G}=-\Omega_{G}$.
\end{definition}

%We notice the following ambiguity in the definition of the potential (\ref{potmon}). 
%This potential, as well as the 2-form 
%(\ref{GolDar}), 
 %is invariant 
%under symplectic transformations acting on the set of cycles $(a_j^-,b_j^-)$. However, $\theta_{mon}$ does change when one modifies some of the  cycles $(a_j^-,b_j^-)$ by a linear combination of cycles $t_j^-$. Under such transformations 
%the coordinates $\zeta_{a_j^-}$ and $\zeta_{b_j^-} $ change by a linear combination of the Casimirs $\log m_j$ with half-integer coefficients.

%The choice of the point $x_k$ at the boundary of each face of $\Sigma^*$ is equivalent to the choice of ciliation
%of the graph $\Sigma$ (recall that the ciliation is the choice of the ordering of edges of the embedded graph 
%at each of its vertices). 
%\begin{figure}[htb]
%\begin{center}
%\includegraphics[width=10cm,angle=0]{Cover_cut.JPG}
%\includegraphics[width=10cm,angle=0]{Figure-6-Covercut.pdf}
%\end{center}
%\caption{\red{Three branch cuts at each zero!!!!}}
%\label{funpol}
%\end{figure}

\subsection{ Graphs $\Sigma_Q$ and $\Sigma^*_Q$ defined by horizontal trajectories of   a quadratic differential}
\la{SSQ}

The construction of the previous section  was purely topological; neither the conformal structure of
the base curve $\CC$ nor its cover $\Ch_\Sigma$ play any role.

Here we specify the  choice of the triangulation $\Sigma$ using a  Gaiotto-Moore-Nietzke differential $Q$.
This triangulation we shall denote by $\Sigma_Q$ and the dual graph by  $\Sigma_Q^*$
(see for details  \cite{IN1,IN2,Alleg,BS}).

Denote the poles of $Q$ by $z_1,\dots,z_n$; they are the vertices of $\Sigma_Q$.
The zeros of  $Q$ are denoted by 
 $x_1,\dots,x_N$ with $N=4g-4+2n$; they are the vertices of $\Sigma_Q^*$. 
 
Topologically, the positions of the  edges of $\Sigma_Q$ and  $\Sigma_Q^*$ are  determined by the horizontal foliation associated 
to $Q$ as defined below.
% notice that in the previous section we did not have any restrictions of  the topology of the graph $\Sigma$ (except that $\Sigma$ is a triangulation of $\CC$). 

 %Recall that in Section \ref{monre1} the topology of the double cover $\Ch_\Sigma$ was determined by 
% the topology of the graph $\Sigma$. 
 %On the other hand, if one starts from the Riemann surface $\CC$ equipped with complex structure and 
 %a chosen quadratic differential $Q$ one can start the construction from defining 
 
 The canonical cover $\Ch_Q$ is defined
 complex-analytically, by the equation (\ref{cancov}):
\be
v^2=Q\;.
\la{cancov1}
\ee 
%and reproduce the graph $\Sigma$ starting from $\Ch_Q$; the genus of $\Ch_Q$ is given by the same formula (\ref{ghdec}); it equals to $\gh=4g-3+n$.
%In this section we show that the covering $\Ch_Q$ is equivalent to the covering $\Ch_{\Sigma_Q}$

%Obviously, it is not sufficient to know only the topology of the cover $\Ch_Q$ to uniquely fix the 
%graph $\Sigma$ such that the covering $\Ch_\Sigma$ topologically coincides with $\Ch_Q$. 
%A can be see from a simple combinatorial counting, for fixed branch points $x_j$ there exist 
%$2^{2g-1}$ inequivalent two-sheeted coverings. On the other hand, the number of triangulations of $\CC$ with $n$ faces is substantially higher; this number can be computed using the machinery of hermitian matrix models (see \cite{LandoZvon} for details). Therefore, to define $\Sigma$ uniquely one needs to use some 
%additional analytical information. This information is encoded in the horizontal foliation defined by the 
%differential $Q$; the resulting graphs turn out to important in the WKB analysis  (see \cite{Alleg,BS}).

Now consider the Abelian integral
of the third kind $v$ on $\Ch$:
\be
z(x)=\int_{x_1}^x v
\la{defco}
\ee
where the initial point of integration can be chosen arbitrarily, but we prefer to choose it to coincide with one 
 the "first" zero $x_1$.

Consider the foliation of $\CC$ (and also of $\Ch_Q$) by horizontal trajectories of $v$:
\be
\Im \int_{x_1}^x v = const\;.
\la{foliation}
\ee
On the cover $\Ch_Q$ this foliation can be oriented in the direction of increase of $\Re  \int_{x_1}^x v $;
since the trajectories on two different sheets of $\Ch_Q$ have different orientation, the projection of such
oriented foliation to $\CC$ is non-oriented.

Assume that $Q$ is generic i.e it does not have any horizontal trajectory connecting two zeros
(this is the definition of  the ``Gaiotto-Moore-Nietzke differential" \cite{GMN}).
% (we notice that this requirement is exactly opposite to the
%Strebel differentials, when all zeros of $Q$ belong to the same horizontal trajectory).
In other words, $Q$ is free from ``saddle connections".

Then  each horizontal trajectory starting at a zero $x_j$ of $Q$ ends at one of the poles $z_k$.
Since three horizontal trajectories meet at each zero, this determines three vertices of the triangle (possibly, self-folded), therefore, defining the triangulation $\Sigma_Q$.
The dual graph with vertices at $x_j$ is denoted by $\Sigma_Q^*$, see Fig.\ref{FigTrian}. 
%The horizontal trajectories connecting zeros with poles then generically split $\CC$ into quadrilaterals; two opposite vertices of each quadrilateral are zeros of $Q$, and two other opposite vertices are poles 
%(some of the opposite vertices can be identified, leading to folded quadrilaterals, see \cite{Alleg,BS}.

%Then the edges of the graph $\Sigma_Q$ are defined to connect the opposite poles of each quadrilateral.
%The edges of the graph $\Sigma_Q^*$ are defined to be diagonals connecting the opposite zeros of each quadrilateral.
%Obviously, the graphs $\Sigma_Q$ and $\Sigma_Q^*$ obtained this way are dual to each other, see Fig.\ref{FigTrian}.
%The graph $\Sigma_Q$ then gives a triangulation of $\CC$ while the graph $\Sigma_Q^*$ is three-valent.
Using the graph  $\Sigma_Q^*$ one defines a  two-sheeted branch covering $\Ch_{\Sigma_Q}$ (as in Section \ref{monre1}) 
by assuming that all edges of $\Sigma_Q^*$ are branch cuts:

\begin{figure}[htb]
\begin{center}
\includegraphics[width=0.5\textwidth]{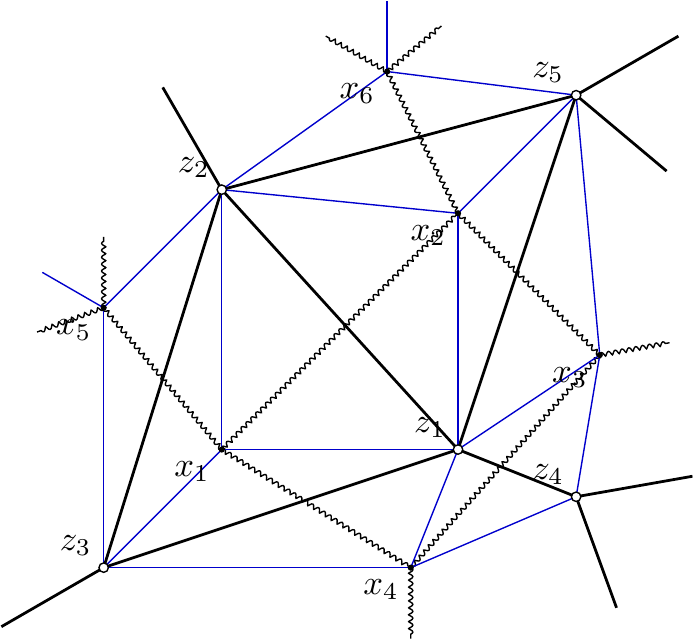}
\end{center}
\caption{Horizontal critical trajectories (blue lines ) connect poles $z_j$ with zeros $x_k$ of $Q$ and form the
critical graph $\Gamma_Q$. Black edges connecting poles $z_j$ form the graph $\Sigma_Q$ (the triangulation of $\CC$) while the  zigzag curves connect zeros $x_j$ and form the dual tri-valent graph $\Sigma_Q^*$.
%Edges of the triangulation $\Sigma_Q$ are shown in blue while the edges of three-valent dual graph $\Sigma_Q^*$ are shown in zigzag.  
All edges of  $\Sigma_Q^*$ are the branch cuts of $\Ch_{\Sigma_Q}$.  }
\label{FigTrian}
\end{figure}

\begin{definition}
Let us denote by $\hat{\Sigma}^*_Q$ the lift of the graph $\Sigma^*_Q$ from $\CC$ to $\Ch$.
Then each face of  $\Sigma^*_Q $ is lifted to the union of two faces of $\hat{\Sigma}^*_Q$
The first sheet $\Ch^{(1)}$ of the canonical cover $\Ch$ is defined to be the union of faces of
$\hat{\Sigma}^*_Q$ which contain the poles $z_j^{(1)}$ with residues $r_j$.
\end{definition}

Now we need to make the following two technical observations
\begin{remark}\rm
\label{remre}
In the distinguished coordinate   $\xi_j(x) := \exp[1/r_j \int_{x_1}^x v] $    (\ref{xijint}) near $z_j^{(1)}$ the differential $v$ writes $v = r_j \d \xi_j/\xi_j$. Then  any horizontal trajectory going to $z_j^{(1)}$ satisfies $\Im( r_j \ln (\xi_j)) =J=$constant. 
Eliminating $\theta = \arg \xi_j$ we thus obtain that, along the horizontal trajectory, the real part satisfies $\Re (r_j \ln (\xi_j) ) = \frac {|r_j|^2}{\Re r_j} \ln |\xi_j| - \frac {\Im r_j}{\Re r_j} J$. Recall that $r_j$ are chosen such that $\Re r_j<0$.  Therefore, $$\Re \int_{x_1}^x v =\Re (r_j \ln (\xi_j) ) \nearrow +\infty $$ 
i.e. it is  increasing as the trajectory approaches the pole $z_j^{(1)}$. Obviously, $\Re \int_{x_1}^x v $ decreases along any horizontal trajectory approaching $z_j^{(2)}$.
%Notice that this feature can be considered as the definition of the first sheet 
%of the canonical double cover.

\end{remark}
\begin{remark}\rm
Consider a zero $x_j$ of $Q$ and any two of the three critical trajectories on the first sheet $\Ch^{(1)}$  going from $x_j$ towards poles $z_k^{(1)}$ and $z_l^{(1)}$. Then the edge of $\Sigma^*_Q$ situated between 
these two critical trajectories is a branch cut of $\Ch$. This observation follows from an elementary local analysis. Namely, in terms of the distinguished local coordinate $\zeta_j=[\int_{x_j}^x v ]^{2/3}$ on $\CC$ we have
$\Re \int_{x_j}^x v = \Re(\zeta_j^{3/2})$. The critical trajectories are defined by $\Im( \zeta^{3/2})=0$;
since we assumed that $\Re(\zeta_j^{3/2})>0$ on each of the three trajectories meeting at $x_j$ (and, moreover, this real part increases along each trajectory) one needs to place a branch cut between any two of them.
  \end{remark}

  The previous two remarks immediately imply the following proposition.

\begin{proposition}\la{ChCh}
The covering of  $\Ch_{\Sigma_Q}$ of $\CC$ obtained  by assuming that {\bf all the edges of $\Sigma_Q^*$
are branch cuts} (as in Section \ref{monre1}) is biholomorphically equivalent to  the canonical covering $\Ch_Q$ defined by (\ref{cancov1}).
\end{proposition}

%In this proposition we give one representative of the configuration of cuts along the edges of $\Sigma_Q^*$
%such that the corresponding double cover is holomorphically equivalent to $\Ch$.
Notice that one can construct many coverings of $\CC$ choosing various configurations of cuts along the edges of $\Sigma_Q^*$. In general, the number of inequivalent coverings of $\CC$ with $4g-4+2n$
branch points equals to $2^{2g}$. Indeed, each covering corresponds to a group homomorphism ${\bf h}$ from
$\pi_1(\CC\setminus \{x_j\}_{j=1}^{4g-4+2n}$ to $S_2$. Choosing the generators of the fundamental group according to (\ref{fgrel}) we see that for any $\delta_j$ we have ${\bf h}(\delta_j)=(1\,\, 2)$.
The permutations ${\bf h}(\a_j)$ and  ${\bf h}(\b_j)$ can be arbitrary; therefore, one has $2^{2g}$ 
inequivalent coverings. All of these coverings can be realized by various choices of branch cuts along the edges of $\Sigma_Q^*$ since the number of edges of $\Sigma_Q^*$ which is equal to $6g-6+3n$ exceeds the number of generators of the above fundamental group, which is equal to $6g+2n-4$.

On the other hand, the set of all branch coverings obtained by various  choices of branch cuts along edges of $\Sigma_Q^*$ are split into equivalence classes.  Each equivalence class contains $2^{n-1}$ coverings 
which can be obtained as follows: take a given configuration of branch cuts and "interchange" some faces of $\Sigma_Q^*$ between the 1st and second sheet.

However, choosing all edges of $\Sigma_Q$ to be branch cuts, and combining the  proposition (\ref{ChCh})
with proposition (\ref{propSigma}) we get the following statement:
\begin{proposition}\la{ChCh1}
For any tri-valent graph $\Sigma^*$ with vertices at $x_j$'s the branch cover $\Ch_\Sigma$ obtained by 
assuming that all edges of $\Sigma^*$ are branch cuts, is biholomorphically equivalent to the canonical cover
$\Ch$ defined by $v^2=Q$.
\end{proposition}

\begin{example}\rm
 Let $g=n=1$.
In this case there are two zeros, $x_1$ and $x_2$  of $Q$ and $\gh=4g-3+n=2$. A special case of the differential $Q$ is given by 
 $$Q = \wp(z) \d z^2$$
 where $\wp$ is the Weierstra\ss--$\wp$ function. 

 Within the fundamental rectangle  the function  $\wp$ has two simple zeroes $x_1$ and $x_2$,
the graph $\Sigma_Q^*$ has three edges and one face. 
Therefore, there exist {\it four inequivalent coverings} obtained by choosing either one of the edges to be a branch cut, or all three edges simultaneously. It is only the latter covering which is holomorphically equivalent to the canonical covering $v^2=Q$. %Therefore, in this case the canonical covering can not be realized by 
 \end{example}


\begin{thebibliography}{}

%\bibitem{Abikoff} W. Abikoff,   {\it The real analytic theory of Teichm\"uller space}, Lecture Notes in Math., {\bf 820} Springer, Berlin, 144 p. (1980)

\bibitem{AbrSteg} M.Abramovitz, I.A.Stegun, {\it Handbook of Mathematical functions}, Applies Mth.Series (1964)

\bibitem{AlBrid}D. Allegretti, T.Bridgeland, {\it The monodromy of meromorphic projective structures},  Trans. Amer. Math. Soc. {\bf 373}  6321-6367 (2020)

\bibitem{Alleg} D. Allegretti, {\it Voros symbols as cluster coordinates},  J. Topol. {\bf 12},  1031-1068 (2019)

%\bibitem{BabBer} Babelon O., Bernard D., Talon M., {\it Introduction to classical integrable systems},  Cambridge University Press (2004)

\bibitem{BeiDri} A.Beilinson, V.Drinfeld, {\it Opers},  	arXiv:math/0501398 

\bibitem{BK1}
M. Bertola, D. Korotkin, {\it Hodge and Prym tau functions, Jenkins-Strebel differentials and combinatorial model of 
${\mathcal M}_{g,n}$},   Comm.Math.Phys. {\bf 378}, 1279 - 1341 (2020)

\bibitem{BK2iso}
M. Bertola, D. Korotkin, {\it Tau-function ang  monodromy symplectomorphism}, arXiv:1910.03370

\bibitem{BK2}  M. Bertola, D. Korotkin,  {\it  Extended Goldman symplectic structure in Fock-Goncharov coordinates},  \href{http://arxiv.org/abs/1910.06744}{arXiv/1910.06744}

\bibitem{BKN} M. Bertola, D. Korotkin, C. Norton, {\it Symplectic geometry of the moduli space of projective structures in homological coordinates}, Invent.Math.  {\bf 210:3 }, 759–814 (2017)

\bibitem{BS} T. Bridgeland, I. Smith,  {\it  Quadratic differentials as stability conditions},  Publ. Math. de l'IHES, {\bf 121} Issue 1, pp 155–278 (2015)

\bibitem{Bri} T. Bridgeland,
{\it Riemann-Hilbert problems from Donaldson-Thomas theory}, 
Invent. Math.  {\bf 216} 69–124 (2019)
 
\bibitem{Chekhov} Chekhov, L., {\it  Symplectic Structures on Teichmüller Spaces $T_{g,s,n}$ and Cluster Algebras}, Proc. Steklov Inst. Math., {\bf  309}, 87-96 (2020)

\bibitem{Chekhov07} Chekhov, L., {\it Lecture Notes on Quantum Teichm\"uller Theory},  	arXiv:0710.2051

\bibitem{CEO} L.Chekhov, B.Eynard, N.Orantin, {\it Free energy topological expansion for the 2-matrix model}, JHEP {\bf 0612} 053 (2006) 

\bibitem{DKMVZ} P. Deift, T. Kriecherbauer, K. T-R. McLaughlin, S. Venakides and X. Zhou, {\it Uniform asymptotics for polynomials orthogonal with respect to varying 
exponential weights and applications to universality questions in random matrix theory}, Commun. Pure Appl. Math.,  {\bf 52} (1999), 1335-1425.
%\bibitem{DouHub}  Douady, A., Hubbard, J. {\it On the Density of Strebel Differentials}, Inventiones Math., {\bf 30} 175-179 (1975)

\bibitem{EO} B.Eynard, N.Orantin,  {\it Invariants of algebraic curves and topological expansion},  Comm. in Number Theory and Physics, {\bf 1}, No. 2, p. 347-452 (2007)

\bibitem{Fay73} J. D. Fay,  {\it Theta-functions on Riemann surfaces},
Lect.Notes in Math.,  {\bf 352} Springer (1973)

\bibitem{Fay92} J. D. Fay, {\it Kernel functions, analytic torsion and moduli spaces}, Memoirs of AMS, No.464 (1992)

%\bibitem{Earle} Earle, C. J.
%{\it On variation of projective structures}. Riemann surfaces and related topics: Proceedings of the 1978 Stony Brook Conference (State Univ. New York, Stony Brook, N.Y., 1978), pp. 87-99,
%Ann. of Math. Stud., 97, Princeton Univ. Press, Princeton, N.J., 1981. 

\bibitem{Fock1} Fock, V.V., {\it Dual Teichm\"uller spaces}, arXiv:dg-ga/9702018 



%\bibitem{GaKaMa} D. Gallo, M.  Kapovich, A.  Marden,  {\it The monodromy groups of Schwarzian equations on closed Riemann surfaces}, 
%Ann. of Math., {\bf 151}  625-704 (2000)

\bibitem{Gold84} W. Goldman,  {\it The symplectic nature of fundamental groups of surfaces}, Adv. in Math. 
{\bf  54}, 200-225 (1984)

%\bibitem{Gold86} W. Goldman,  {\it Invariant functions on Lie groups and Hamiltonian flows
%of surface group representations}, Invent. math. {\bf 85}, 263-302 (1986)

%\bibitem{Gold09}  W. Goldman,  {\it Trace coordinates on Fricke spaces of some
%simple hyperbolic surfaces},  Handbook of Teichm\"uller theory. Vol. II, 611-684, IRMA Lect. Math. Theor. Phys., 13, Eur. Math. Soc., Zorich, (2009)
 
 \bibitem{GMN} D.Gaiotto, G.Moore, A.Neitzke, {\it Wall-crossing, Hitchin Systems, and the WKB Approximation}
 arXiv:0807.4723 
 
 \bibitem{GT} D.Gaiotto, J.Teschner, {\it Irregular singularities in Liouville theory }, JHEP {\bf 2012} (2012)

%
\bibitem{HawSch}  N. S. Hawley, M.   Schiffer, {\it Half-order differentials on Riemann surfaces}, Acta Math. {\bf 115}  199-236 (1966)

\bibitem{IN1}
 K.Iwaki, T.Nakanishi,
{\it Exact WKB analysis and cluster algebras},  J. Phys. A: Math. Theor. 47 (2014) 474009 

\bibitem{IN2} K.Iwaki, T.Nakanishi,
{\it Exact WKB analysis and cluster algebras II: Simple poles, orbifold points, and generalized cluster algebras},
 Int. Math. Res. Not. {\bf 2016} 4375-4417 (2016) 
 
%\bibitem{Hejhal} Hejhal, D.  A.
%{\it Monodromy groups and linearly polymorphic functions}.
%Acta Math. {\bf 135}  no. 1, 1-55 (1975)

\bibitem{KT05} T.Kawai, Y.Takei, {\it Algebraic analysis of Singular perturbation theory}, Transl. of Math. Monographs, vol.227, AMS (2005)

%\bibitem{Hubbard} Hubbard, J. H. {\it The monodromy of projective structures}, in "Riemann surfaces and Related Topics", Proceedings of 1978 Stony Brook conference, Princeton Univ.Press (1980)

\bibitem{KalKor} C. Kalla, D. Korotkin, {\it Baker-Akhiezer kernel and tau-functions on moduli spaces of meromorphic differentials}, Commun. Math.Physics. {\bf 331} Iss. 3 1191-1235 (2014)

%\bibitem{Kenyon} R.Kenyon, {\it Lectures on dimers}, https://www.math.brown.edu/~rkenyon/papers/dimerlecturenotes.pdf

%\bibitem{Kapov} M. Kapovich,  {\it On monodromy of complex projective structures}, Invent. Math. {\bf 119} 243-265 (1995)

%\bibitem{Klimov} R.Klimov, {\it On higher order terms in WKB expansion of Yang-Yang function}, to appear

%\bibitem{Kawai} Kawai, S. {\it The symplectic nature of the space of projective connections on Riemann surfaces}, Math Ann {\bf 305} 161-182 (1996)

%\bibitem{KalKor} C.Kalla, D.Korotkin, {\it Baker-Akhiezer kernel and tau-functions on moduli spaces of meromorphic differentials}, Communications in Mathematical Physics, {\bf 331} Issue 3 1191-1235 (2014)

\bibitem{Knizhnik} Knizhnik V. G., {\it Analytic fields on Riemann surfaces II}, Comm. Math. Phys. {\bf 112} 567–590 (1987)

\bibitem{HP} Knizhnik V.G., {\it  Multiloop amplitudes in the theory of quantum strings and complex geometry}, Sov. Phys. Uspekhi {\bf 32} (11), 945–971 (1989)

\bibitem{Kon}  M. Kontsevich,  {\it Intersection theory on the moduli space of curves and the matrix Airy function}, Comm. Math. Phys. {\bf 147}  no. 1, 1–23
(1992)



\bibitem{JDG} A.Kokotov, D.Korotkin, {\it Tau-functions on spaces of Abelian
differentials and higher genus generalization of Ray-Singer formula},
J. Diff. Geom. {\bf 82}  35-100 (2009)


\bibitem{Kor}
D. Korotkin, {\it Periods of meromorphic quadratic differentials and Goldman bracket},  in "Topological recursion and its influence in analysis, geometry, and topology" : 2016 AMS von Neumann Symposium, 
AMS  Proceedings of Symposia in Pure Mathematics, ed. by M.L.Chiu-Chu and M.Mulase, p. 491-516

\bibitem{KorRev}
D. Korotkin, "Bergman tau function: from Einstein equations and Dubrovin-Frobenius manifolds to geometry of moduli spaces",
in  "Integrable Systems and Algebraic Geometry", ed. by R.Donagi and T.Shaska,  LMS Lecture Note Series,   Cambridge University Press, in press, arXiv:1812.03514

%\bibitem{KSZ}
 %D.Korotkin, A.Sauvaget, P.Zograf, {\it Tau functions, Prym-Tyurin classes and loci of degenerate differentials},
 %Math.Annalen, {\bf 375} p. 213-246 (2019)

%\bibitem{MRL}  D. Korotkin, P.   Zograf, {\it Tau function and moduli of differentials}, Math.Res.Lett., {\bf 18} No.3, 447-458  (2011)



\bibitem{contemp}  D. Korotkin,  P. Zograf, {\it Tau function and the Prym class}, Contemporary Mathematics, {\bf 593} 241-261 (2013)

%\bibitem{LandoZvon}    S.Lando, A.Zvonkin,    {\it     Graphs on Surfaces and Their Applications}, Springer, 455 pp (2004)



\bibitem{NRS} N. Nekrasov, A.  Rosly, S.  Shatashvili, {\it Darboux coordinates, Yang-Yang functional, and gauge theory}, arXiv: 1103.3919 [hep-th]

\bibitem{Quillen} D.Quillen, {\it Determinants of Cauchy-Riemann operators over a Riemann surface}, 
Func. Anal. and Its Appl. {\bf 19} 31-34 (1985)

\bibitem{Thurston} W. P.Thurston, {\it The Geometry and Topology of 3-Manifolds}, Princeton University Notes (1980).
%\bibitem{Voros}  A. Voros, {\it The return of the quartic oscillator. The complex WKB method}, Ann.Inst. Henri Poincar\'e  {\bf 39}   211–338 (1983)

\bibitem{Wasow} Wasow, W., {\it Asymptotic Expansions for Ordinary Differential Equations},  1987

%\bibitem{SeiWit} Seiberg, N.,  Witten, E., {\it Monopole Condensation and Confinement In N=2 Supersymmetric Yang-Mills Theory}, Nucl. Phys. {\bf B426} 19-52 (1994),  hep-th/9407087 







\end{thebibliography}
\end{document}